\documentclass[twocolumn]{aastex631}

\newcommand\dustpy{\texttt{DustPy}}

\shorttitle{\texttt{DustPy}: A Python Package for Dust Evolution in Protoplanetary Disks}
\shortauthors{Stammler \& Birnstiel 2022}
\graphicspath{{./}{figures/}}

\usepackage[T1]{fontenc}
\usepackage[english]{babel}

\usepackage[fleqn]{amsmath}
\usepackage{amssymb}
\usepackage{bbold}

\begin{document}

\title{\texttt{DustPy}: A Python Package for Dust Evolution in Protoplanetary Disks}

\author[0000-0002-1589-1796]{Sebastian Markus Stammler}
\affiliation{University Observatory, Faculty of Physics, Ludwig-Maximilians-Universität München, Scheinerstr. 1, D-81679, Munich, Germany}

\author[0000-0002-1899-8783]{Tilman Birnstiel}
\affiliation{University Observatory, Faculty of Physics, Ludwig-Maximilians-Universität München, Scheinerstr. 1, D-81679, Munich, Germany}
\affiliation{Exzellenzcluster ORIGINS, Boltzmannstr. 2, D-85748 Garching, Germany}



\begin{abstract}

    Many processes during the evolution of protoplanetary disks and during planet formation are highly sensitive to the sizes of dust particles that are present in the disk: The efficiency of dust accretion in the disk and volatile transport on dust particles, gravoturbulent instabilities leading to the formation of planetesimals, or the accretion of pebbles onto large planetary embryos to form giant planets are typical examples of processes that depend on the sizes of the dust particles involved. Furthermore, radiative properties like absorption or scattering opacities depend on the particle sizes. To interpret observations of dust in protoplanetary disks, a proper estimate of the dust particle sizes is needed.

    We present \dustpy{}: a \texttt{Python} package to simulate dust evolution in protoplanetary disks. \dustpy{} solves gas and dust transport including viscous advection and diffusion as well as collisional growth of dust particles. \dustpy{} is written with a modular concept, such that every aspect of the model can be easily modified or extended to allow for a multitude of research opportunities.

\end{abstract}



\section{Introduction} \label{sec:intro}

Dust plays an important role in many processes of planet formation. Interstellar micrometer-sized dust particles accumulate in protoplanetary disks due to angular momentum conservation and grow to millimeter-sized pebbles via collisional growth. However, various growth barriers prevent particles from growing directly into planetesimals, e.g., the charge barrier, the bouncing barrier, the drift barrier, or the fragmentation barrier. As soon as particles reach millimeter sizes gravoturbulent instabilities can play an important role. These instabilities have the ability to concentrate particles in pebble clouds, which can subsequently collapse under their own gravity into planetesimals if they are massive enough \citep[see][]{youdin2005ApJ...620..459Y, johansen2007Natur.448.1022J}. These planetesimals can collide to form larger bodies and eventually planetary embryos. Leftover dust pebbles in the disk accrete onto these embryos to assemble terrestrial planets or the cores of giant planets \citep[see][]{ormel2017ASSL..445..197O}.

The efficiency of said processes like planetesimal formation \citep{krapp2019ApJ...878L..30K, paardekooper2020MNRAS.499.4223P} or pebble accretion onto planetary embryos \citep{liu2018A&A...615A.138L, ormel2018A&A...615A.178O} is highly sensitive to the sizes of dust particles available in protoplanetary disks. To understand the formation of planets and to interpret the population and composition of observed exoplanets it is therefore crucial to know which particle sizes can exist at any specific time and location in the lifetime of protoplanetary disks. Furthermore, it is important to understand the size evolution of dust particles when interpreting observations of dust in protoplanetary disks \citep{sierra2021ApJS..257...14S}.

To simulate the collisional evolution of micrometer-sized dust particles up to planets it is unfeasible to simulate every dust particle individually. Several techniques have been developed in the past to overcome this. One is the Monte Carlo method wherein several dust particles are combined into a few representative particles, whose evolution can be simulated \citep[see][]{ormel2007A&A...461..215O, zsom2008A&A...489..931Z, drazkowska2013A&A...556A..37D}. Another method is to simulate the evolution of a particle size distribution instead of individual particles \citep[see][]{weidenschilling1980Icar...44..172W, nakagawa1981Icar...45..517N, dullemond2005A&A...434..971D, brauer2008A&A...480..859B, birnstiel2010A&A...513A..79B}. The advantage of Monte Carlo methods is that it is relatively easy to include additional particle properties like electrical charge, porosity, or composition, while this is a rather complex task in the case of particle distributions \citep[see][]{okuzumi2009ApJ...707.1247O, stammler2017A&A...600A.140S}. Monte Carlo methods, however, are computationally expensive, while methods with particle distributions can cover longer time spans of disk evolution.

We developed the \texttt{Python} package \dustpy{}\footnote{\dustpy{} repository: \href{https://github.com/stammler/dustpy}{https://github.com/stammler/dustpy}\\Documentation: \href{https://stammler.github.io/dustpy/}{https://stammler.github.io/dustpy/}\\Installation of the latest version: \texttt{pip install dustpy}\\\dustpy{} v1.0.1: \href{https://doi.org/10.5281/zenodo.6874878}{https://doi.org/10.5281/zenodo.6874878}}, which simulates the evolutions of a dust mass distribution in protoplanetary disks accounting for collisional growth and transport of dust particles, as well as the evolution of the gas disk. It can be used to simulate the evolution of the gas and dust within a protoplanetary disk over its entire life span.

The main object of \dustpy{} is to calculate the evolution of the gas surface density $\Sigma_\mathrm{g}$ and $N_m$ dust surface densities $\Sigma_{\mathrm{d},i}$ of different particle masses in a protoplanetary disk with $N_r$ radial grid cells, including viscous evolution of the gas, advection and diffusion of the dust, as well as collisional dust growth by solving the Smoluchowski equation. It is therefore one-dimensional in space. \dustpy{} itself uses the \texttt{Simframe} framework for scientific simulations \citep{stammler2022JOSS....7.3882S}, which allows the user to easily customize every aspect of the model or to extend it with additional functionality. This publication, therefore, discusses the default functionality of \dustpy{}, i.e., the model that is run without any customization. \dustpy{} is written in \texttt{Python} to allow for easy access and great flexibility. Computationally expensive routines are written in \texttt{Fortran}. However, the main focus is on usability, not on optimizing the execution time.

This publication is structured as follows. Section \ref{sec:gas_evo} introduces the relevant equations for gas evolution. In section \ref{sec:dust_evo} we discuss dust evolution, which consists of two major parts: dust transport in section \ref{sec:dust_trans} and collisional dust growth in section \ref{sec:dust_growth}. In section \ref{sec:bench} we compare the coagulation algorithm of \dustpy{} to test cases that have analytical solutions. In section \ref{sec:examples} we present example simulations that show the full potential of \dustpy{} customizations. Finally, in section \ref{sec:summary} we summarize the features and caveats of \dustpy{}.

\section{Gas Evolution} \label{sec:gas_evo}

\dustpy{} assumes an axisymmetric disk. All quantities, therefore, only have one spatial coordinate, the radial distance $r$ from the star. If applicable, \dustpy{} assumes vertical hydrostatic equilibrium in the $z$ direction, i.e., the height above the midplane of the disk. By default, \dustpy{} viscously evolves the gas surface density via the viscous advection-diffusion equation
\begin{equation}
    \label{eqn:gas_evo}
    \frac{\partial}{\partial t} \Sigma_\mathrm{g} + \frac{1}{r} \frac{\partial}{\partial r} \left( r \Sigma_\mathrm{g} v_\mathrm{g}\right) = S_\mathrm{ext}
\end{equation}
including external source terms $S_\mathrm{ext}$, which can be used to implement, for example, infall of matter onto the disk or losses due to photoevaporation. In the default model $S_\mathrm{ext}$ is set to zero. The radial gas velocity is given by
\begin{equation}
    \label{eqn:v_gas}
    v_\mathrm{g} = Av_\mathrm{visc} +2B\eta v_\mathrm{K},
\end{equation}
with the Keplerian velocity $v_\mathrm{K} = \sqrt{\frac{GM_*}{r}}$ and the viscous accretion velocity given by \citet{lynden-bell1974MNRAS.168..603L} as
\begin{equation}
    \label{eqn:gas_vvisc}
    v_\mathrm{visc} = - \frac{3}{\Sigma_\mathrm{g}\sqrt{r}} \frac{\partial}{\partial r} \left(\Sigma_\mathrm{g}\nu\sqrt{r}\right),
\end{equation}
with the kinematic viscosity $\nu$. $G$ is the gravitational constant and $M_*$ the mass of the central star. In the default \dustpy{} model the kinematic viscosity is given by $\nu=\alpha \frac{c_\mathrm{s}^2}{\Omega_\mathrm{K}}$, with the $\alpha$ viscosity parameter introduced by \citet{shakura1973A&A....24..337S}, the sound speed $c_\mathrm{s}$ and the Keplerian frequency $\Omega_\mathrm{K}=\sqrt{\frac{GM_*}{r^3}}$. $\eta$ is the pressure gradient parameter given by
\begin{equation}
    \eta = -\frac{1}{2}\left(\frac{H_\mathrm{P}}{r}\right)^2\frac{\partial\log P}{\partial\log r},
\end{equation}
with the pressure scale height $H_\mathrm{P} = \frac{c_\mathrm{s}}{\Omega_\mathrm{K}}$. The parameters $A$ and $B$ in equation (\ref{eqn:v_gas}) are used to implement the dynamic back reaction of dust particles onto the gas. In the default model \dustpy{} uses $A=1$ and $B=0$, i.e., no back reaction. This is accurate as long as the dust mass is small compared to the gas mass. With decreasing $A$ and increasing $B$, gas accretion can be halted or even reversed. \citet{garate2020A&A...635A.149G} implemented back reaction of dust particles onto the gas in a region with an increased dust-to-gas ratio caused by a "traffic jam" at the water-ice line into \dustpy{}. For details on the implementation of $A$ and $B$ we refer to that publication.

\subsection{Algorithm} \label{sec:gas_evo_algo}

\begin{figure}[tb]
    \centering
    \includegraphics[width=\linewidth]{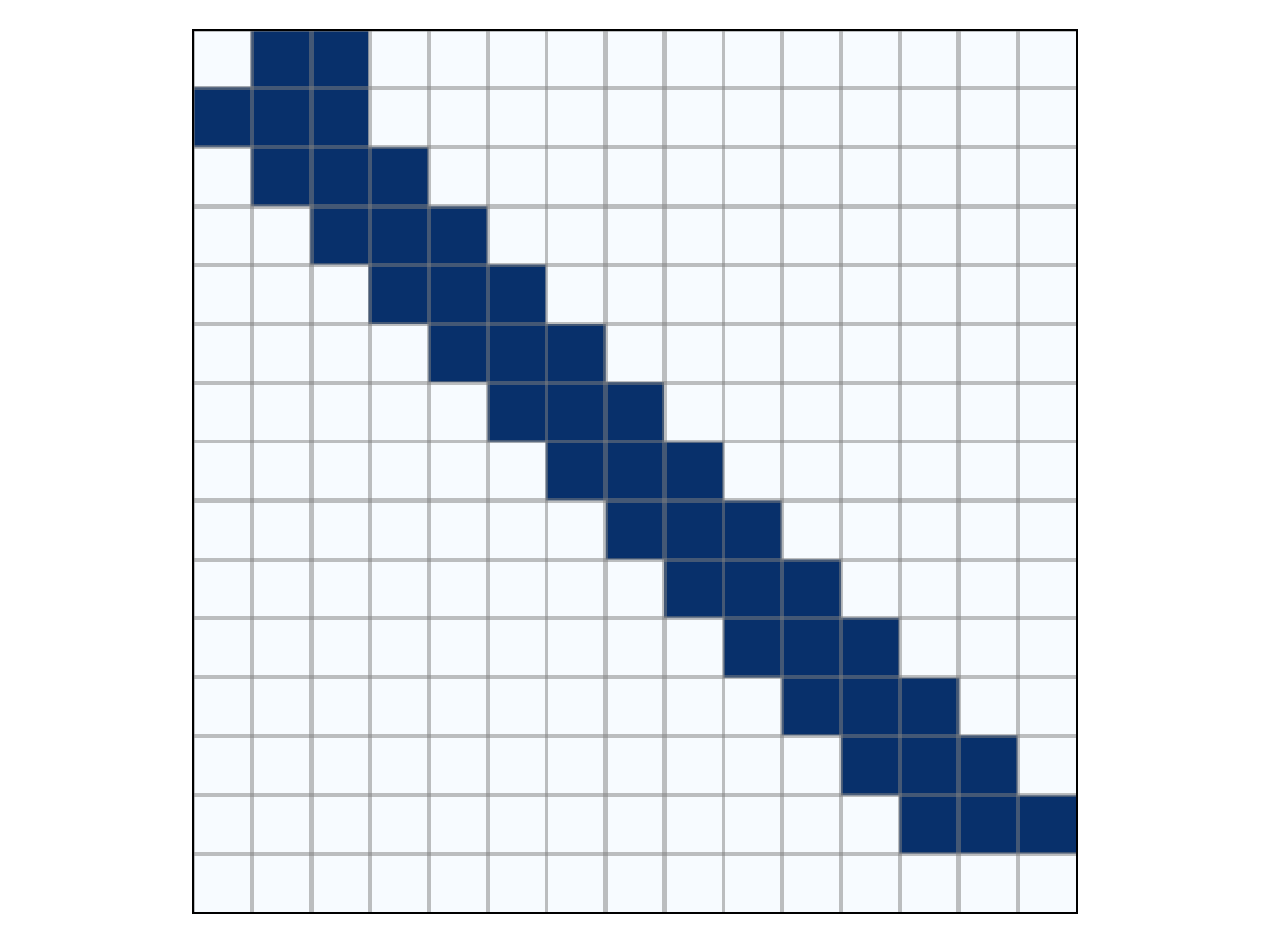}
    \caption{Sketch of the gas Jacobian in an example with 15 radial grid cells. The Jacobian consists of the main diagonal and one upper and one lower off-diagonal, because the gas can only interact with itself or with neighboring grid cells. Exceptions are the first and the last rows, which are used to set the boundary conditions.}
    \label{fig:gas_jacobian}
\end{figure}

Equation (\ref{eqn:gas_evo}) can be discretized and written as a matrix equation implicitly in $\vec{\Sigma}_\mathrm{gas}^n$, where the superscript $n$ represents the time coordinate
\begin{equation}
    \label{eqn:gas_matrix}
    \frac{\vec{\Sigma}_\mathrm{g}^{n+1}-\vec{\Sigma}_\mathrm{g}^n}{\Delta t} = \mathbb{J}\cdot \vec{\Sigma}_\mathrm{g}^{n+1} + \vec{S}_\mathrm{ext}.
\end{equation}
The Jacobian $\mathbb{J}$ is a tri-diagonal matrix, since the radial grid cells only interact with themselves or with neighboring grid cells.

A sketch of the Jacobian $\mathbb{J}$ can be seen in \autoref{fig:gas_jacobian}. Exceptions to the tri-diagonal shape of the Jacobian are the first and the last rows, which are used to set the boundary conditions. By default the inner boundary is set to a constant gradient, while the outer boundary is set to the gas floor value to prevent inflow of gas through the outer boundary. Since most of the elements of $\mathbb{J}$ are zero, \dustpy{} uses the \texttt{scipy.sparse} package \citep{virtanen2020NatMe..17..261V} to store the Jacobian in a sparse matrix format.

Equation (\ref{eqn:gas_matrix}) can be solved for $\vec{\Sigma}_\mathrm{g}^{n+1}$ via
\begin{equation}
    \vec{\Sigma}_\mathrm{g}^{n+1} = \left(\mathbb{1}-\Delta t\mathbb{J}\right)^{-1} \cdot \left( \vec{\Sigma}_\mathrm{g}^n + \Delta t \vec{S}_\mathrm{ext}^n \right)
\end{equation}
by inverting the matrix $\mathbb{1}-\Delta t\mathbb{J}$. To achieve this, the matrix is factorized with \texttt{scipy.sparse.linalg.splu}, before solving the system of equations with \texttt{scipy.sparse.linalg.SuperLU.solve}.

\section{Dust Evolution} \label{sec:dust_evo}

Dust evolution in \dustpy{} consists of two parts: dust transport and dust growth. Dust transport is calculated by solving the advection-diffusion equation similar to gas evolution. Dust growth is calculated by solving the Smoluchowski equation. The dust quantities consist of $N_m$ different dust species of different masses at every radial position.

\subsection{Dust Transport} \label{sec:dust_trans}

To account for dust transport \dustpy{} solves the advection-diffusion equation \citep{clarke1988MNRAS.235..365C} for every dust species $i$
\begin{equation}
    \label{eqn:dust_adv_diff}
    \begin{split}
        &\frac{\partial}{\partial t} \Sigma_{\mathrm{d}, i} + \frac{1}{r}\frac{\partial}{\partial r} \left( r\Sigma_{\mathrm{d}, i} v_{\mathrm{d}, i} \right) \\
        &\quad- \frac{1}{r} \frac{\partial}{\partial r} \left[ r D_i \Sigma_\mathrm{g} \frac{\partial}{\partial r} \left( \frac{\Sigma_{\mathrm{d}, i}}{\Sigma_\mathrm{g}} \right) \right] = S_{\mathrm{ext}, i}.
    \end{split}
\end{equation}
The radial dust velocity $v_{\mathrm{d}, i}$ is given by
\begin{equation}
    \label{eqn:dust_v_rad}
    v_{\mathrm{d}, i} = \left( v_\mathrm{g} + 2v_\mathrm{drift}^\mathrm{max}\mathrm{St}_i\right) \frac{1}{\mathrm{St}_i^2+1},
\end{equation}
with the maximum drift velocity given by
\begin{equation}
    v_\mathrm{drift}^\mathrm{max} = \frac{1}{2}Bv_\mathrm{visc} - A\eta v_\mathrm{K},
\end{equation}
where $A$ and $B$ are the back reaction coefficients introduced in \autoref{eqn:v_gas}. The Stokes number $\mathrm{St}_i$ is a measure of the aerodynamic size of a dust particle. \dustpy{} considers by default two aerodynamic drag regimes: the Epstein and the Stokes I regimes
\begin{equation}
    \mathrm{St}_i =
    \begin{cases}
        \frac{\pi}{2} \frac{a_i \rho_{\mathrm{s}, i}}{\Sigma_\mathrm{g}}                      & \text{for $a_i < \frac{9}{4}\lambda_\mathrm{mfp}$ (Epstein)} \\
        \frac{2\pi}{9} \frac{a_i^2\rho_{\mathrm{s},i}}{\lambda_\mathrm{mfp}\Sigma_\mathrm{g}} & \text{else (Stokes I)}
    \end{cases}
\end{equation}
with the dust particle radius $a_i$, the dust bulk density $\rho_{\mathrm{s},i}$, and the mean free path of the gas $\lambda_\mathrm{mfp}$. The Stokes~I regime is typically only important for large particles in the inner parts of protoplanetary disks. The dust diffusivity in equation (\ref{eqn:dust_adv_diff}) is taken from \citet{youdin2007Icar..192..588Y} and is given by
\begin{equation}
    \label{eqn:dust_diff}
    D_i = \frac{\delta_\mathrm{r}c_\mathrm{s}^2}{\Omega_\mathrm{K}} \frac{1}{1+\mathrm{St}_i^2}.
\end{equation}
The parameter $\delta_\mathrm{r}$ describes the strength of radial diffusion of the dust particles and is similar to the turbulent $\alpha$ parameter.

\subsection{Dust Growth} \label{sec:dust_growth}

Modeling collisional dust growth in protoplanetary disks is challenging. To form an Earth-like planet out of micrometer-sized dust particles one would need to simulate the evolution of about $10^{40}$ individual dust particles, which is not feasible. To overcome this problem a number of strategies have been developed in the past. One is the so-called Monte Carlo method, in which many physical particles are combined into a few representative particles, whose evolution can be calculated \citep[see][]{ormel2007A&A...461..215O, zsom2008A&A...489..931Z, drazkowska2013A&A...556A..37D}.

\texttt{DustPy}, on the other hand, calculates dust growth by solving the Smoluchowski equation

\begin{equation}
    \label{eqn:smolu}
    \begin{split}
        \frac{\partial}{\partial t} n\left( m \right) = \int\limits_0^\infty \int\limits_0^{m'} & K\left(m, m', m''\right) R\left(m', m''\right) \quad \times \\
        \times \quad & n\left(m'\right) n\left(m''\right) \mathrm{d}m'' \mathrm{d}m' \\
        -n\left(m\right) \int\limits_0^{\infty} & R\left(m, m'\right)n\left(m'\right)\mathrm{d}m'.
    \end{split}
\end{equation}
Instead of tracking individual particles, \texttt{DustPy} calculates the collisional evolution of a distribution $n\left(m\right)$ of particles with masses $m$. The first double integral on the right-hand side sums over all possible collisions of particles with masses $m'$ and $m''$ and collision rate $R\left(m',m''\right)$. The matrix $K\left(m, m', m''\right)$ holds information about the collision outcomes of each collision and describes the amount that gets added into $n\left(m\right)$ from a single collision of particles with masses $m'$ and $m''$. A perfectly sticking collision would be described with $K\left(m, m', m''\right) = \delta\left( m - \left( m' + m'' \right) \right)$.

The upper boundary of the inner integral is $m'$ instead of $\infty$, because collisions of particles with masses $m'$ and $m''$ are identical to collisions of particles with masses $m''$ and $m'$ and should not be counted twice. The negative term on the right-hand side accounts for the particles that get removed from the distribution, because they have collided with other particles.

\texttt{DustPy} discretizes $n\left(m\right)$ on a mass grid with $N_m$ mass bins by integrating $n\left(m\right)$ over the mass bin width
\begin{equation}
    \label{eqn:dust_discretization}
    n_i = \int\limits_{m_{i-\frac{1}{2}}}^{m_{i+\frac{1}{2}}} n\left(m\right) \mathrm{d}m.
\end{equation}
With this the discretized Smoluchwoski equation can be written as
\begin{equation}
    \frac{\partial}{\partial t} n_k = \sum\limits_{i=1}^{N_m} \sum\limits_{j=1}^i K_{ijk} R_{ij} n_i n_j - n_k \sum\limits_{j=1}^{N_m} n_j R_{jk} \left( 1 + \delta_{jk}\right).
    \label{eqn:smolu_disc}
\end{equation}
Please note the Kronecker $\delta$ in the second term on the right-hand side. For equal particle collisions $\left(j=k\right)$ two particles from the same mass bin have to be removed from the distribution.

This section discusses the various challenges in implementing equation (\ref{eqn:smolu_disc}) into the numerical algorithm of \dustpy{}. This is rather technical. Readers that want to skip the derivation can continue reading at section \ref{sec:coll_rates}. In a typical \dustpy{} simulation the user does not need to modify the collisional subroutines, unless the goal is to implement a custom collision model.

\subsubsection{Coagulation} \label{sec:dust_coag}

The case of pure coagulation, i.e., perfect sticking of two particles forming a new larger body, has a number of computational challenges. Firstly, the mass grid of \texttt{DustPy} is logarithmically spaced to cover a large dynamic range from submicrometer-sized particles to large boulders. This has the disadvantage that the resulting mass of two colliding particles $m^\mathrm{coll} = m_i+m_j$ will in general not fall exactly onto the mass grid itself, but in between two mass bins. We follow the approach of \citet{brauer2008A&A...480..859B} to linearly distribute the newly formed particle between the two adjacent mass bins. Assuming the mass of the particle resulting from a sticking collision of particle $m_i$ and $m_j$ falls in between the two mass bins $m_m \leq m^\mathrm{coll} < m_n$, we split the coagulation rate linearly between both mass bins

\begin{equation}
    \label{eqn:lin_cont}
    K_{ijk} =
    \begin{cases}
        \epsilon     & \text{if $m_k = m_m$} \\
        1 - \epsilon & \text{if $m_k = m_n$} \\
        0            & \text{else},
    \end{cases}
\end{equation}
where $\epsilon$ is given by

\begin{equation}
    \label{eqn:eps_coag}
    \epsilon = \frac{m_n - m^\mathrm{coll}}{m_n - m_m}.
\end{equation}
If $m^\mathrm{coll} = m_m \Rightarrow \epsilon = 1$ and the entire particle will be distributed into mass bin $m_m$. Since by definition $m_n > m^\mathrm{coll}$, mass will always be distributed into a mass bin that is larger than the combined mass of the colliding particles leading to artificial growth. The mass grid, therefore, needs to be fine enough to limit this numerical inaccuracy. This limitation is discussed in section \ref{sec:bench_coag} in more detail.

\begin{figure*}[tb]
    \centering
    \includegraphics[width=\linewidth]{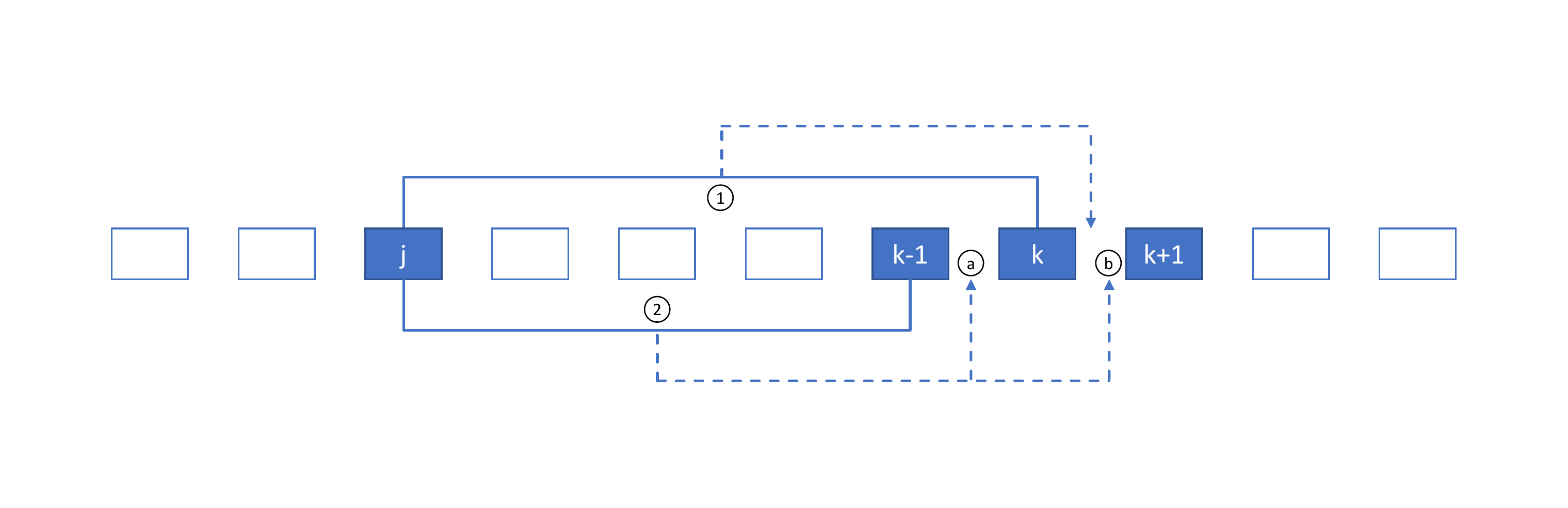}
    \caption{Sketch of the mass grid with three special types of collisions highlighted. Type 1: particle of mass $m_j$ collides with particle of mass $m_k$ with the resulting mass being between $m_k$ and $m_{k+1}$. Type 2a: particle of mass $m_j$ collides with particle of mass $m_{k-1}$ with the resulting mass being between $m_{k-1}$ and $m_{k}$. Type 2b: particle of mass $m_j$ collides with particle of mass $m_{k-1}$ with the resulting mass being between $m_{k}$ and $m_{k+1}$.}
    \label{fig:sketch}
\end{figure*}

Another challenge is purely computational and is caused by the limited precision of computers. For double-precision numbers, for example, we face the problem that $m_i + m_j = m_i$, if $m_i$ is more than $15$ orders of magnitudes more massive than $m_j$. This would prevent large particles from growing by sweeping up many small particles. This problem can be solved by rearranging the sums in the discrete Smoluchowski equation. We, again, follow the approach of \citet{brauer2008A&A...480..859B}. \autoref{fig:sketch} shows a sketch of three types of particle collision that are dealt with separately in this section.

Starting from equation (\ref{eqn:smolu_disc}) we can separate the diagonals from the first term on the right-hand side, i.e., the equal-mass particle collisions, and combine it with the Kronecker $\delta$ of the negative term
\begin{equation}
    \begin{split}
        \dot{n}_k^\mathrm{s} = &\sum\limits_{i=1}^{k} R_{ii}^\mathrm{s} n_i^2 \left( K_{iik} - \delta_{ik} \right) \\
        + &\sum\limits_{i=1}^{k} \sum\limits_{j=1}^{i-1} K_{ijk} R_{ij}^\mathrm{s} n_i n_j  - n_k &\sum\limits_{j=1}^{N_m} n_j R_{jk}^\mathrm{s}.
    \end{split}
    \label{eqn:smolu_diag_split}
\end{equation}
The superscript "s" denotes that these are the source terms and collisions rates for purely sticking collisions. The sum over $i$ does not need to go all the way to $N_m$, but only up to $k$, because sticking collisions involving particles larger than $m_k$ can never positively contribute to $n_k$. We now look more closely at the second and third terms on the right-hand side and separate the case $k=i$ from the second term, i.e., those collisions that can be affected by machine-precision errors, when a large particle with mass $m_k$ is sweeping up a small particle $m_j$ such that the resulting mass is in between $m_k$ and $m_{k+1}$
\begin{equation}
    \begin{split}
        & \sum\limits_{i=1}^{k} \sum\limits_{j=1}^{i-1} K_{ijk} R_{ij}^\mathrm{s} n_i n_j - n_k \sum\limits_{j=1}^{N_m} n_j R_{jk}^\mathrm{s} \\
        = \quad & \sum\limits_{i=1}^{k-1} \sum\limits_{j=1}^{i-1} K_{ijk} R_{ij}^\mathrm{s} n_i n_j \\
        + &\sum\limits_{j=1}^{k-1} K_{kjk} R_{kj}^\mathrm{s} n_k n_j - n_k \sum\limits_{j=1}^{N_m} n_j R_{jk}^\mathrm{s}.
    \end{split}
\end{equation}
The second term on the right-hand side with $K_{kjk}$ represents these special collisions for which particles with masses $m_k$ and $m_j$ collide, but still have a positive contribution to $n_k$ (type 1 in \autoref{fig:sketch}). The term only describes particle collisions for which $m_k + m_j \leq m_{k+1}$, otherwise the resulting mass of the collision would be too large to positively contribute to $n_k$. We therefore introduce a number $c$, which is defined as the smallest integer for which the condition $m_k + m_{k+1-c} \leq m_{k+1}$ is fulfilled. In general $c$ would depend on $k$. But since the mass grid of \texttt{DustPy} is regular logarithmic, $c$ will be a constant as long as the mass grid does not change. We can now replace the upper boundary of the sum in the second term with $k+1-c$ and combine it with the respective negative part of the sum in the third term
\begin{equation}
    \begin{split}
        & \sum\limits_{j=1}^{k-1} K_{kjk} R_{kj}^\mathrm{s} n_k n_j - n_k \sum\limits_{j=1}^{N_m} n_j R_{jk}^\mathrm{s} \\
        = \quad & \sum\limits_{j=1}^{k+1-c} R_{jk}^\mathrm{s} n_j n_k \left( K_{kjk} - 1 \right) \\
        - n_k &\sum\limits_{j=k+2-c}^{N_m} n_j R_{jk}^\mathrm{s}.
    \end{split}
\end{equation}
Please note that $c\ge2$, since the equal-size collisions $\left( j=k \right)$ are already included in the first term in equation \ref{eqn:smolu_diag_split}. In any case, mass grids with $K_{kkk} \ne 0$ that lead to $m_k \le m_k + m_k < m_{k+1}$ would have fewer than $\log_2 10 \approx 3.3$ mass bins per decade. Simulations should have at least seven mass bins per decade for simple collision models \citep{Ohtsuki1990}, and even more for complex collision models \citep{drazkowska2014A&A...567A..38D}. Further, note that the collision rates are symmetric, i.e.,  $R_{jk}^\mathrm{s} = R_{kj}^\mathrm{s}$. Collisions of particle $m_j$ with $m_k$ occur at the same rate as collisions of particle $m_k$ with $m_j$. Since $K_{kjk}$ in the first term means that $m_k \leq m_k+m_j < m_{k+1}$ we can set $m_m = m_k$ and use equation (\ref{eqn:lin_cont}) to get
\begin{equation}
    \begin{split}
        K_{kjk} - 1 = \epsilon - 1 &= \frac{m_{k+1} - \left( m_k + m_j \right)}{m_{k+1} - m_k} - 1 \\
        &= - \frac{m_j}{m_{k+1} - m_k}.
    \end{split}
\end{equation}
In collisions prone to machine-precision errors, a computer would falsely calculate $m_k + m_j = m_k$. Already manipulating $K_{kjk} - 1$ in advance in these collisions eliminates these errors. Using this for the first term and combining it with the second term we get
\begin{equation}
    \begin{split}
        & \sum\limits_{j=1}^{k+1-c} \left( K_{kjk} - 1 \right) R_{jk}^\mathrm{s} n_j n_k - n_k \sum\limits_{j=k+2-c}^{N_m} n_j R_{jk}^\mathrm{s} \\
        = \quad & \sum\limits_{j=1}^{N_m} D_{jk} R_{jk}^\mathrm{s} n_j n_k,
    \end{split}
\end{equation}
with
\begin{equation}
    D_{jk} =
    \begin{cases}
        - \frac{m_j}{m_{k+1} - m_k} & \text{if $j \leq k+1-c$} \\
        - 1                         & \text{if $j > k+1-c$}
    \end{cases}
\end{equation}
Up to this point the coagulation equation reads
\begin{equation}
    \begin{split}
        \dot{n}_k^\mathrm{s} = &\sum\limits_{i=1}^{k} R_{ii}^\mathrm{s} n_i^2 \left( K_{iik} - \delta_{ik} \right) + \sum\limits_{i=1}^{k-1} \sum\limits_{j=1}^{i-1} K_{ijk} R_{ij}^\mathrm{s} n_i n_j \\
        + &\sum\limits_{j=1}^{N_m} D_{jk} R_{jk}^\mathrm{s} n_j n_k.
    \end{split}
\end{equation}
Now we look at the second term on the right-hand side for the case $i = k-1$
\begin{equation}
    \sum\limits_{j=1}^{k-2} K_{k-1,jk}\ R_{k-1,j}^\mathrm{s}\ n_{k-1}\ n_j.
\end{equation}
These are the other types of collision that can be affected by machine-precision errors. In this case particles with masses $m_{k-1}$ and $m_j$ collide and have a positive contribution to $n_k$. We can distinguish two cases here. In the first case the resulting mass of the colliding particles falls in between $m_{k-1} \leq m_{k-1} + m_j < m_k$ (type 2a in \autoref{fig:sketch}). This means $m_k = m_n$ in equation (\ref{eqn:lin_cont}) and therefore
\begin{equation}
    \begin{split}
        K_{k-1,jk} = 1 - \epsilon & = 1 - \frac{m_k - \left( m_{k-1} + m_j \right)}{m_k - m_{k-1}} \\
        & = \frac{m_j}{m_k - m_{k-1}}.
    \end{split}
\end{equation}
These collisions are identical to type 1 but look at the mass that is distributed into the larger mass bin. In the second case the resulting mass falls in between $m_k \leq m_{k-1} + m_j < m_{k+1}$ (type 2b in \autoref{fig:sketch}). Here we have $m_k = m_m$ in equation (\ref{eqn:lin_cont}) and therefore
\begin{equation}
    \begin{split}
        &K_{k-1,jk } = \epsilon \\
        &= \frac{m_{k+1} - \left( m_{k-1} + m_j \right)}{m_{k+1}-m_k} \Theta \left( m_{k+1} - m_{k-1} - m_j \right) \\
        &= \left[ 1 - \frac{m_j + m_{k-1} - m_k}{m_{k+1}-m_k} \right] \Theta \left( m_{k+1} - m_{k-1} - m_j \right).
    \end{split}
\end{equation}
Cases with $m_{k-1} + m_j >\geq m_{k+1}$ do not contribute positively toward $n_k$. The Heaviside step function takes care of these cases. In both cases either $1-\epsilon$ or $\epsilon$ itself can be affected by machine-precision errors. It is therefore advisable to manipulate directly these cases in advance as shown above. We can now split the sum into both cases using the constant $c$ that has been introduced earlier
\begin{equation}
    \begin{split}
        & \sum\limits_{j=1}^{k-2} K_{k-1,jk}\ R_{k-1,j}^\mathrm{s}\ n_{k-1}\ n_j \\
        = \quad & \sum\limits_{j=1}^{k-2} E_{jk}\ R_{k-1,j}^\mathrm{s}\ n_{k-1}\ n_j
    \end{split}
\end{equation}
The matrix $E$ is given by
\begin{equation}
    E_{jk} =
    \begin{cases}
        \frac{m_j}{m_k - m_{k-1}}                                                                                      & \text{if $j \leq k-c$} \\
        \left[ 1 - \frac{m_j - m_{k-1} - m_{k}}{m_{k+1}-m_{k}} \right] \Theta \left( m_{k+1} - m_{j} - m_{k-1} \right) & \text{if $j>k-c$}.
    \end{cases}
\end{equation}
The full coagulation equation now reads
\begin{equation}
    \begin{split}
        \dot{n}_k^\mathrm{s} = & \sum\limits_{i=1}^k \left( K_{iik} -\delta_{ik} \right) R_{ii}^\mathrm{s}n_i^2 + \sum\limits_{j=1}^{N_m} D_{jk} R_{jk}^\mathrm{s} n_j n_k \\
        + &\sum\limits_{i=1}^{k-2} \sum\limits_{j=1}^{i-1} K_{ijk} R_{ij}^\mathrm{s} n_i n_j + \sum\limits_{j=1}^{k-2} E_{jk}\ R_{k-1,j}^\mathrm{s}\ n_{k-1}\ n_j.
    \end{split}
\end{equation}
It is useful to bring the equation into a double sum form
\begin{equation}
    \label{eqn:double_sum}
    \begin{split}
        \dot{n}_k^\mathrm{s} = & \sum\limits_{i=1}^{N_m} \sum\limits_{j=1}^{N_m} \left(K_{ijk}-\delta_{ik}\right)R_{ij}^\mathrm{s} n_i n_j \delta_{ij} + \sum\limits_{i=1}^{N_m} \sum\limits_{j=1}^{N_m} D_{ij} R_{ij}^\mathrm{s} n_i n_j \delta_{jk} \\
        + &\sum\limits_{i=1}^{N_m} \sum\limits_{j=1}^{N_m} K_{ijk} R_{ij}^\mathrm{s} n_i n_j \Theta\left( k - i - \frac{3}{2} \right) \Theta\left( i - j - \frac{1}{2} \right) \\
        + & \sum\limits_{i=1}^{N_m} \sum\limits_{j=1}^{N_m} E_{j,i+1}R_{ij}^\mathrm{s} n_{i}n_j \delta_{i, k-1}\Theta\left(k-j-\frac{3}{2}\right),
    \end{split}
\end{equation}
where the Kronecker $\delta$ and the Heaviside step function $\Theta$ are used to pick the correct values and ranges for $i$ and $j$. In that way the coagulation equation can be written with one single double sum. Note that the inner sum has to go up until $N_m$, because the matrices $D_{ji}$ and $E_{jk}$ are not symmetric. Since $i$ and $j$ are integer number, the terms of $\frac{1}{2}$ and $\frac{3}{2}$ in the Heaviside step functions are used to avoid a potentially undefined behavior for $\Theta\left(0\right)$.

From a computational perspective it is beneficial to bring the equation into a symmetrical form to save half of the iterations. \texttt{DustPy}, therefore, solves the following equation
\begin{equation}
    \label{eqn:coag_sticking}
    \dot{n}_k^\mathrm{s} = \sum\limits_{i=1}^{N_m} \sum\limits_{j=1}^i C_{ijk}R_{ij}^\mathrm{s}n_in_j,
\end{equation}
with
\begin{equation}
    \label{eqn:Cijk}
    C_{ijk} = \begin{cases}
        \tilde{C}_{ijk} + \tilde{C}_{jik} & \text{if $i\geq j$} \\
        0                                 & \text{else}
    \end{cases}
\end{equation}
and
\begin{equation}
    \label{eqn:Ctildeijk}
    \begin{split}
        \tilde{C}_{ijk} = &\frac{1}{2}K_{ijk}\delta_{ij} + D_{ij}\delta_{jk} \\
        + &K_{ijk}\Theta\left(k-i-\frac{3}{2}\right)\Theta\left(i-j-\frac{1}{2}\right) \\
        + & E_{j,i+1}\delta_{i, k-1}\Theta\left(k-j-\frac{3}{2}\right).
    \end{split}
\end{equation}
Please note that diagonal entries $\left(i=j\right)$ must not be counted twice. All diagonals need therefore a factor of $\frac{1}{2}$. Only the first two terms of equation (\ref{eqn:double_sum}) contain diagonals. The first term contains only diagonal entries, while the second term contains both diagonal and off-diagonal entries. However, since $c \ge 2$, this means for the diagonals of the second term $D_{kk} = -1$. This is one particle that gets removed from the distribution in collisions with two equal-size particles. The other particle is removed by $\delta_{ik}$ in the first term of equation (\ref{eqn:double_sum}). We can therefore omit the $\delta_{ik}$ in the first term and the factor of $\frac{1}{2}$ for the second term in equation (\ref{eqn:Ctildeijk}).

For any given collision of particles with masses $m_i$ and $m_j$, $C_{ijk}$ will either have zero, three, or four nonzero elements. $C_{ijk}$ will always have two positive entries for $k=m$ and $k=n$ (the two mass bins in between which the resulting collisional mass falls) and negative entries for $k=i$ and $k=j$ (the colliding particles), leading to four nonzero entries. In special cases, when $i=j$, $k=i$, or $k=j$, this number can be reduced to three. If $m_i + m_j$ is larger than the largest mass of the mass grid $C_{ijk}$ will be set to zero to prevent mass loss through the upper boundary of the mass grid.

A peculiar property of $C_{ijk}$ for any particle collision is
\begin{equation}
    \sum\limits_{k=1}^{N_m} C_{ijk} =
    \begin{cases}
        -1 \\
        0 \qquad \text{if $m_i + m_j > m_{N_m}$}.
    \end{cases}
\end{equation}
Since the coagulation equation described above works on number densities, the sum of $C_{ijk}$ over $k$ for any combination of $i$ and $j$ has to be $-1$ as long as the mass of the colliding particles is within the mass grid. Two particles collide, stick, and form a single larger particle. Therefore, for every sticking collision the total number of particles is reduced by one.

$C_{ijk}$ only needs to be calculated once in the beginning of the simulation as long as the mass grid does not change. Because there are a maximum of four nonzero elements for any combination of $i$ and $j$, the coagulation problem is of the order $\mathcal{O}\left(N_m^2\right)$.

Equation (\ref{eqn:coag_sticking}) can be written in matrix form
\begin{equation}
    \frac{\partial}{\partial t} \vec{n} = \mathbb{J}^s \cdot \vec{n}
\end{equation}
with the sticking Jacobian $J^\mathrm{s}$ being defined as
\begin{equation}
    J^\mathrm{s}_{ki} = \sum\limits_{j=1}^i C_{ijk} R_{ij} n_j.
\end{equation}

\begin{figure*}[tb]
    \centering
    \includegraphics[width=\textwidth]{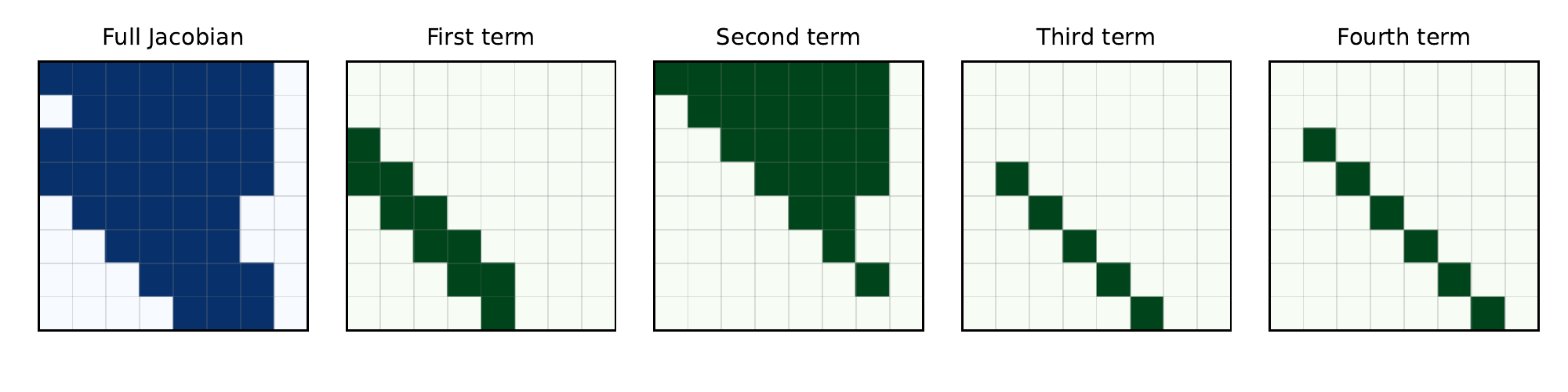}
    \caption{Sketch of the sticking Jacobian with the contributions of the four terms in equation (\ref{eqn:Ctildeijk}).}
    \label{fig:J_comp}
\end{figure*}

A sketch of the structure of the Jacobian with the contributions of the four terms in the definition of $\tilde{C}$ in equation (\ref{eqn:Ctildeijk}) is shown in \autoref{fig:J_comp} for a mass resolution of seven mass bins per decade. The last column is always empty, since collisions with particles of mass $m_{N_m}$ will always result in a particle exceeding the mass grid. In this setup the element $J_{21}$ is empty, because it represents collisions involving at least one particle of mass $m_1$ that have a positive contribution to $n_2$. However, the mass grid is fine enough, such that $m_1 + m_1 > m_3$, which means that $n_2$ cannot be filled from these types of collisions. The first term represents equal particle collisions, the second term contains the $\mathbb{D}$-matrix with the negative contributions to the distribution, the fourth term contains the contribution of the $\mathbb{E}$-matrix, and the third term contains the remaining collisions.

\subsubsection{Fragmentation} \label{sec:dust_frag}

If the relative velocity of the colliding particles exceeds the fragmentation velocity, particles fragment rather than stick and grow. \texttt{DustPy} distinguishes by default two types of fragmentation events: full fragmentation and erosion.

Full fragmentation means that both colliding particles fully fragment, leaving behind a fragment distribution that follows a power law:
\begin{equation}
    \label{eqn:fragment_dist}
    n\left(m\right)\mathrm{d}m \propto m^\gamma \mathrm{d}m.
\end{equation}
The exponent $\gamma$ has to be determined experimentally. \texttt{DustPy} uses by default $\gamma = -\frac{11}{6}$, taken from \citet{dohnanyi1969JGR....74.2531D}. Erosion, on the other hand, happens when both colliding particles differ significantly in mass. The smaller projectile particle then fully fragments while chipping off some mass from the larger target particle. The outcome of a erosive collision is a fragment distribution and a slightly less massive remnant target particle. In \texttt{DustPy} the transition between full fragmentation and erosion is by default at a particle mass ratio of $10$.

To calculate the contribution of fragmenting collisions, \texttt{DustPy} uses the $\mathcal{O}\left(N_m^2\right)$ algorithm developed by \citet{rafokov2020ApJS..247...65R}. We slightly modified the algorithm to make it strictly mass conserving and to account for the \texttt{DustPy} code units, where the dust quantities are integrated over the mass bin.
For this purpose we define a normalized fragment distribution:
\begin{equation}
    \varphi_{ij} = \frac{m_j^{1+\gamma}}{\sum\limits_{j=1}^i m_j^{1+\gamma}} \Theta\left( i-j+\frac{1}{2} \right) \times \frac{1}{\mathrm{g\,cm^3}}.
\end{equation}
$\varphi_{ij}$ is the amount that gets added to $n_j$ from a fragment distribution with a total mass $1$ and a largest fragment mass of $m_i$. The Heaviside step function sets $\varphi_{ij}$ to zero if the index $j$ is greater than the largest mass bin of the fragment distribution. The exponent is $1+\gamma$ instead of $\gamma$, because the quantity is integrated over the mass bin. As another quantity we define the total mass of fragments that is created in a single particle collision event:
\begin{equation}
    \label{eqn:mass_normalization}
    A_{ij} =
    \begin{cases}
        m_i + m_j                & \text{for full fragmentation} \\
        \left(1+\chi \right) m_j & \text{for erosion}.
    \end{cases}
\end{equation}
For fully fragmenting collisions the fragment mass is the total mass of the colliding particles. For erosive collisions the projectile particle chips off a fraction of $\chi$ of its own mass from the target particle. In \texttt{DustPy} $\chi=1$ by default. Without loss of generality, $m_j$ is always the mass of the smaller projectile particle. Another quantity that is needed is the largest mass, $m_k$, of the fragment distribution. The index of the largest fragment is given by
\begin{equation}
    k_{ij}^\mathrm{lf} =
    \begin{cases}
        i & \text{for full fragmentation} \\
        j & \text{for erosion}
    \end{cases}
\end{equation}
In fully fragmenting collisions the fragment distribution goes all the way up to the largest particle. In erosive collisions, the largest particle of the fragment distribution has the mass of the projectile particle. With these quantities we can now sum up the contribution of all collisions to the total fragment distribution multiplied with their individual fragment mass and weighted by their collision rates and store them in a vector, $A_k^*$, at the position of the largest fragment:
\begin{equation}
    A^*_k = \sum\limits_{i=1}^{N_m} \sum\limits_{j=1}^i A_{ij} R_{ij}^\mathrm{f} n_i n_j \delta_{k,k_{ij}^\mathrm{lf}},
\end{equation}
where the superscript "f" denotes the collision rates for fragmenting collisions. The contribution from fragments of all collisions into $n_k$ is then given by
\begin{equation}
    \dot{n}_k^\mathrm{fragments} = \sum\limits_{i=k}^{N_m} A_i^* \varphi_{ik}.
\end{equation}
In both cases, full fragmentation and erosion, the smaller projectile particle will fully fragment and has to be removed from the particle distribution:
\begin{equation}
    \dot{n}_k^\mathrm{projectile} = - \sum\limits_{i=1}^{N_m} \sum\limits_{j=1}^{i} R_{ij}^\mathrm{f} n_i n_j \delta_{jk}.
\end{equation}
Similarly, the larger target particle has to be removed in fully fragmenting collisions. In erosive collisions, however, the target particle has to be removed from the distribution and then added, as remnant particle, at another place in the distribution.
\begin{equation}
    \begin{split}
        \dot{n}_k^\mathrm{target} = -&\sum\limits_{i=1}^{N_m}\sum\limits_{j=1}^{i}R_{ij}^\mathrm{f}n_in_j\delta_{ik} \\
        + &\sum\limits_{i=1}^{N_m}\sum\limits_{j=1}^i\tilde{H}_{ijk}R_{ij}^\mathrm{f}n_in_j
    \end{split}
\end{equation}
The matrix $\tilde{H}_{ijk}$ is similar to $K_{ijk}$ in equation (\ref{eqn:lin_cont}) from the previous section and decides between which two mass bins the remnant particle has to be distributed, but with the mass of the remnant particle instead of the total mass of both collision partners. If the remnant particle has a mass $m_{i-1} \leq m_i - \chi m_j<m_i$, mass would be removed and then added into the target particle's mass bin $n_i$. If that is the case, a similar manipulation as for the coagulation can be performed to avoid machine-precision errors. Since the mass grid of \texttt{DustPy} is logarithmically spaced and the transition between full fragmentation and erosion is defined by the mass ratio of the colliding particles, we can define a constant $p$, such that full fragmentation happens if $j \ge i-p$
\begin{equation}
    \dot{n}_k^\mathrm{target} = \sum\limits_{i=1}^{N_m}\sum\limits_{j=1}^{i-p-1}H_{ijk}R_{ij}^\mathrm{f}n_in_j -\sum\limits_{i=1}^{N_m}\sum\limits_{j=i-p}^{i}R_{ij}^\mathrm{f}n_in_j\delta_{ik}.
\end{equation}
The first term now holds both the positive and the negative contribution for the target particle from erosive collisions. We can distinguish two cases. In the first case the positive and negative contributions can be combined, since they both affect the same mass bin:

1. $m_{i-1} \leq m_i - \chi m_j < m_i$:
\begin{equation}
    \begin{alignedat}{3}
        &H_{ij,i-1} &&= \epsilon = \frac{m_i-\left(m_i - \chi m_j\right)}{m_i-m_{i-1}} &&= \frac{\chi m_j}{m_i-m_{i-1}} \\
        &H_{iji} &&= 1-\epsilon-1 = -\epsilon &&= - \frac{\chi m_j}{m_i-m_{i-1}}
    \end{alignedat}
\end{equation}

2. $m_{m} \leq m_i-\chi m_j < m_n < m_i$:
\begin{equation}
    \begin{alignedat}{4}
        &H_{ij,n-1} &&= \epsilon &&= &&\frac{m_n-\left(m_i - \chi m_j\right)}{m_n-m_{m}} \\
        &H_{ij,n} &&= 1-\epsilon &&= &&1-\frac{m_n-\left(m_i - \chi m_j\right)}{m_n-m_{m}} \\
        &H_{iji} && &&= &&-1
    \end{alignedat}
\end{equation}
The full equation for fragmentation and erosion is therefore the sum of all three contributions:
\begin{equation}
    \dot{n}_k^\mathrm{f} = \dot{n}_k^\mathrm{fragments} + \dot{n}_k^\mathrm{projectile} + \dot{n}_k^\mathrm{target}
\end{equation}

\begin{figure}[tb]
    \centering
    \includegraphics[width=\linewidth]{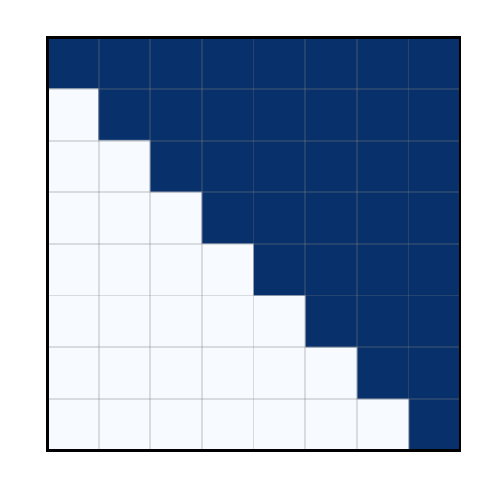}
    \caption{Sketch of the fragmentation Jacobian.}
    \label{fig:J_frag}
\end{figure}

The source terms of fragmentation can also be written in matrix form. A sketch of the fragmentation Jacobian is shown in \autoref{fig:J_frag} for a model where every collision leads to a fragmentation event. The fragmentation Jacobian is a simple upper triangular matrix.

This fragmentation and erosion prescription is of the order $\mathcal{O}\left(N_m^2\right)$. The full dust growth equation can then be simply combined to
\begin{equation}
    \frac{\partial}{\partial t} n_k = \dot{n}_k^\mathrm{s} + \dot{n}_k^\mathrm{f}
\end{equation}
with the dust Jacobian being the sum of the sticking and fragmentation Jacobians:
\begin{equation}
    \mathbb{J}^\mathrm{coag} = \mathbb{J}^\mathrm{s} + \mathbb{J}^\mathrm{f}.
\end{equation}

\subsubsection{Collision Rates} \label{sec:coll_rates}

The collision rates for sticking/fragmenting collisions are the product of the geometrical cross section, the relative velocities of the particles, and the sticking/fragmentation probabilities, respectively:
\begin{equation}
    R^\mathrm{s/f} = \frac{1}{1+\delta_{ij}} \sigma_\mathrm{geo} v_{ij}^\mathrm{rel}p^\mathrm{s/f}
    \label{eqn:coll_rates}
\end{equation}
with $\sigma_\mathrm{geo} = \pi \left( a_i + a_j \right)^2$.
However, please note that the \dustpy{} quantities are vertically integrated, which has been ignored so far. Vertical integration of the Smoluchowski equation introduces a correction factor, which will be incorporated into the collision rates. This is discussed in section \ref{sec:vert_int} in more detail. Further note, that a population of $N$ equal-sized particles has $\frac{1}{2}N\left(N-1\right)$ possible collisions amongst each other, which is $\approx \frac{N^2}{2}$ in the limit of large $N$. Equal particle collisions, therefore, need a factor of $\frac{1}{2}$, which is incorporated via the $\delta_{ij}$ in the collision rates in equation (\ref{eqn:coll_rates}).

\subsubsection{Relative Velocities} \label{sec:dust_vrel}

\begin{figure*}[tb]
    \centering
    \includegraphics[width=\textwidth]{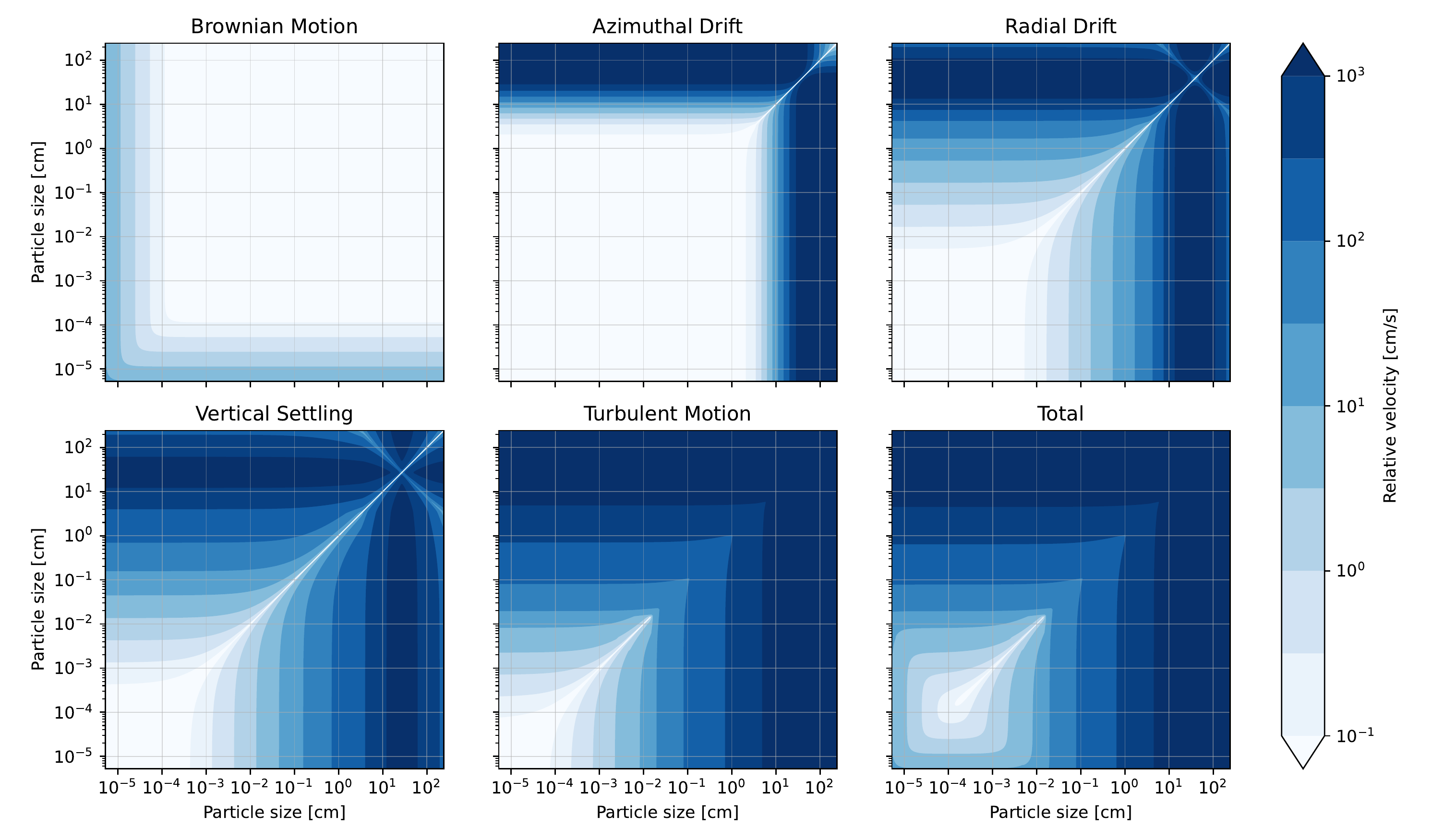}
    \caption{Example of the different source of relative velocities in the default model of \dustpy{} at a distance of 1\,AU.}
    \label{fig:rel_vel}
\end{figure*}

\dustpy{} considers by default five different sources of relative velocities between dust particles: Brownian motion, radial and azimuthal drift, vertical settling, and turbulence.

\autoref{fig:rel_vel} shows all five contributions to the relative velocities in an example simulation of the default \dustpy{} model at a distance of 1\,AU from the star. Brownian motion is especially important as a driver of initial dust growth, when the particles are rather small.

\paragraph{Brownian motion}

The relative velocities due to Brownian motion are given by
\begin{equation}
    v_{ij}^\mathrm{rel,\,brown} = \min \left( \sqrt{\frac{8k_\mathrm{B}T\left(m_i+m_j\right)}{\pi m_im_j}}, \ c_\mathrm{s} \right).
\end{equation}
Since this formula is diverging for very small particle masses, the relative velocities are limited to the sound speed $c_\mathrm{s}$. However, one should note that for very small particles and high temperatures, the relative velocities due to Brownian motion can easily exceed typical values for the fragmentation velocity. In the simple collision model that is used by default in \dustpy{}, there is no distinction on particle size when deciding between sticking and fragmentation. Even though these small particles would in reality still stick (or bounce) at these velocities \citep{chokshi1993ApJ...407..806C, blum2008ARA&A..46...21B}, \dustpy{} would treat those collisions as fragmentation events.

\paragraph{Azimuthal drift}

Since dust particles of different sizes have different degrees of sub-Keplerian motion, this leads to a relative velocity in azimuthal direction, which is given by
\begin{equation}
    v_{ij}^\mathrm{rel,\,azi} = \left| v_\mathrm{drift}^\mathrm{max} \left( \frac{1}{1+\mathrm{St}_i^2} - \frac{1}{1+\mathrm{St}_j^2} \right) \right|.
\end{equation}
Dust particles of the same Stokes number do not experience any relative velocity due to azimuthal drift, because they drift at the same speed.

\paragraph{Radial drift}

Dust particles of different sizes have different radial drift speeds. This induces relative velocities between dust particles. They are given by
\begin{equation}
    v_{ij}^\mathrm{rel,\,rad} = \left| v_{\mathrm{d},i} - v_{\mathrm{d},j} \right|
\end{equation}
with the radial dust velocities from equation (\ref{eqn:dust_v_rad}).

\paragraph{Vertical settling}

Dust particles of different sizes settle with different velocities toward the midplane. \dustpy{} uses the descriptions of \citet{dullemond2004A&A...421.1075D} and \citet{birnstiel2010A&A...513A..79B} to account for this effect:
\begin{equation}
    v_{ij}^\mathrm{rel,\,sett} = \left| h_i\min\left(\mathrm{St}_i, \frac{1}{2}\right) - h_j\min\left(\mathrm{St}_j, \frac{1}{2}\right) \right| \Omega_\mathrm{K},
\end{equation}
where $h_i$ is the dust scale height, given by \citet{dubrulle1995Icar..114..237D} as
\begin{equation}
    \label{eqn:dust_h}
    h_i = H_\mathrm{P} \frac{\delta_z}{\delta_z+\mathrm{St}_i}.
\end{equation}
$\delta_z$ is the vertical settling parameter similar to the turbulent $\alpha$ parameter.

\paragraph{Turbulent motion}

To calculate the relative velocities due to turbulent motion we follow the prescription of \citet{ormel2007A&A...466..413O}. Instead of the turbulent $\alpha$ parameter we use the $\delta_t$ parameter, which works in the identical way, but allows us to disentangle both effects.

The total relative velocity is then the quadratic sum of all contributions
\begin{equation}
    v_{ij}^\mathrm{rel} = \sqrt{ \sum\limits_{k} \left( v_{ij}^{\mathrm{rel, }k} \right)^2 }
\end{equation}

\subsubsection{Coagulation/Fragmentation Probabilities} \label{sec:dust_prob}

\begin{figure*}[tb]
    \centering
    \includegraphics[width=\textwidth]{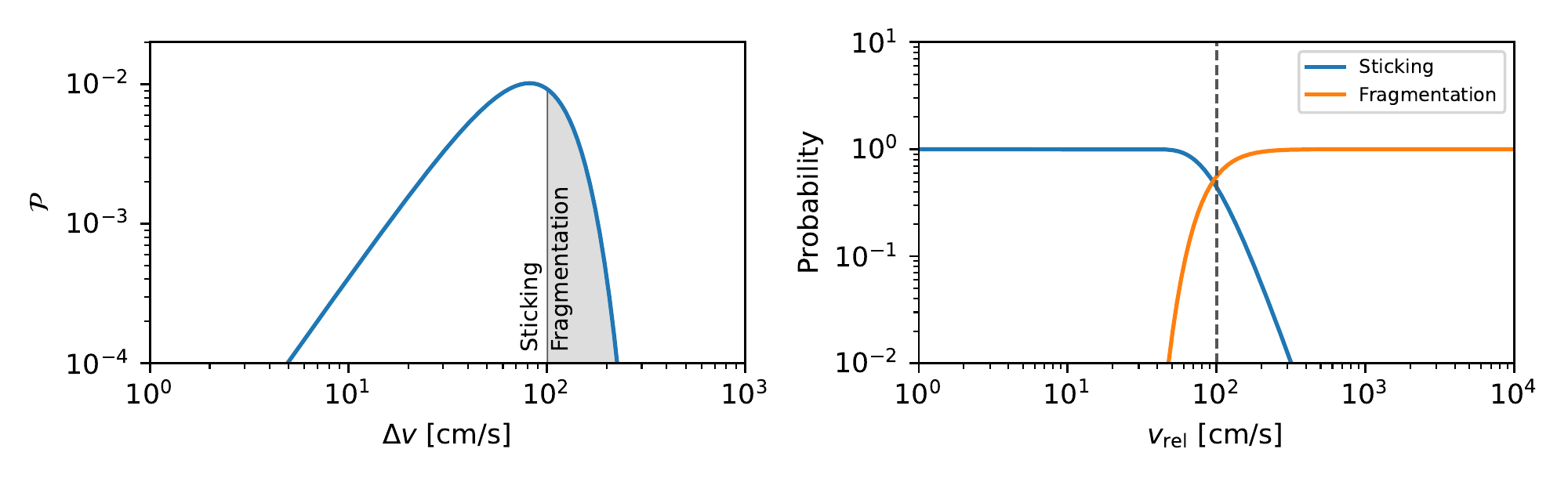}
    \caption{\emph{Left}: Maxwell-Boltzmann velocity distribution for a RMS velocity of $1\,\frac{\mathrm{m}}{\mathrm{s}}$. Assuming a fragmentation velocity of $1\,\frac{\mathrm{m}}{\mathrm{s}}$ some of the particles may fragment while some others can still grow. \emph{Right}: sticking and fragmentation probabilities depending on the RMS velocity assuming a Maxwell-Boltzmann velocity distribution and a fragmentation velocity of $1\,\frac{\mathrm{m}}{\mathrm{s}}$.}
    \label{fig:velocity_distribution}
\end{figure*}

If the relative collision velocity exceeds the fragmentation velocity, particles start to fragment instead of growing to larger bodies. In the default \dustpy{} model the fragmentation velocity is set to $v_\mathrm{frag} = 1\,\mathrm{m}\mathrm{s}^{-1}$.

Different particles, however, do not collide with a single relative velocity as described in section \ref{sec:dust_vrel}. Instead, the relative velocities follow the Maxwell-Boltzmann distribution:
\begin{equation}
    \mathcal{P}\left(\Delta v; v_\mathrm{rms}\right) = \sqrt{\frac{54}{\pi}} \frac{\Delta v^2}{v_\mathrm{rms}^3} \exp \left[ -\frac{3}{2} \left( \frac{\Delta v}{v_\mathrm{rms}} \right)^2 \right],
\end{equation}
with the root mean square velocity $v_\mathrm{rms}$, which \dustpy{} assumes to be the single velocity derived in section \ref{sec:dust_vrel}.

The collision rate for fragmenting collisions described in section \ref{sec:coll_rates} would therefore be an integral over all possible relative velocities in the Maxwell-Boltzmann distribution, which are above the fragmentation velocity:
\begin{equation}
    \begin{split}
        R_{ij}^\mathrm{f} &= \int\limits_{v_\mathrm{frag}}^\infty \sigma_\mathrm{geo} \Delta v \mathcal{P}\left(\Delta v, v_\mathrm{rms}\right) \mathrm{d}\Delta v.
    \end{split}
\end{equation}
This integral has an analytical solution and therefore the fragmentation probability in equation (\ref{eqn:coll_rates}) can be written as
\begin{equation}
    \label{eqn:p_frag}
    \begin{split}
        p_{ij}^\mathrm{f} &= \frac{R_{ij}^\mathrm{f}}{\sigma_\mathrm{geo}\Delta\bar{v}} \\
        &= \left( \frac{3}{2} \left( \frac{v_\mathrm{frag}}{v_{ij}^\mathrm{rel}}\right)^2+1 \right) \exp \left[ -\frac{3}{2} \left( \frac{v_\mathrm{frag}}{v_{ij}^\mathrm{rel}} \right)^2 \right],
    \end{split}
\end{equation}
with the mean velocity of the Maxwell-Boltzmann distribution $\Delta \bar{v} = \sqrt{\frac{8\pi}{3}} v_{ij}^\mathrm{rel}$. In that way, it is sufficient to only calculate one relative velocity per particle collision while accounting for the velocity distribution in the fragmentation probability.

The sticking probability is then given by
\begin{equation}
    p_{ij}^\mathrm{s} = 1-p_{ij}^\mathrm{f}.
\end{equation}
Bouncing, i.e., neither sticking nor fragmentation, is not included in the default model of \texttt{DustPy}, but can be easily implemented if $p_{ij}^\mathrm{s}+p_{ij}^\mathrm{f}<1$. \autoref{fig:velocity_distribution} shows the Maxwell-Boltzmann distribution in the case of an RMS velocity equal to the fragmentation velocity and the sticking/fragmentation probabilities for different relative velocities.

The same approach of a velocity distribution was used by \citet{windmark2012A&A...544L..16W}, while \citet{birnstiel2010A&A...513A..79B} originally used a simple formula for the transition between sticking and fragmentation at the fragmentation velocity.

\subsubsection{Vertical Integration} \label{sec:vert_int}

So far we have ignored the vertical dimension of the disk. The Smoluchowski equation discussed in previous chapters works on volume densities. \dustpy{}, on the other hand, only has one spatial dimension, the distance from the star $r$. Here we describe the method of \citet{birnstiel2010A&A...513A..79B} in vertically integrating the Smoluchowski equation. We assume that the vertical dust distribution can be described with a Gaussian:
\begin{equation}
    n_k\left(r, z\right) = n_k\left(r, 0\right) \cdot \exp\left[-\frac{1}{2}\left(\frac{z}{h_k\left(r\right)}\right)^2\right],
\end{equation}
with the dust scale height $h_k$ given by equation (\ref{eqn:dust_h}). Integration over $z$ leads to
\begin{equation}
    \begin{split}
        N_k\left(r\right) &= \int\limits_{-\infty}^\infty n_k\left(r, z\right) \mathrm{d}z \\
        &= \sqrt{2\pi} h_k\left(r\right) n_k\left(r, 0\right)
    \end{split}
\end{equation}

Every single term in the collisional dust source terms introduced in sections \ref{sec:dust_coag} and \ref{sec:dust_frag} contain the product of two densities with the collision rates $R_{ij} n_i n_j$. Integrating this term over $z$ and assuming to zeroth order that the collision rates do not depend on $z$ leads to
\begin{equation}
    \begin{split}
        & R_{ij} \int\limits_{-\infty}^{\infty} n_i \left(r,z\right) n_j \left(r,z\right) \mathrm{d}z \\
        = \quad &R_{ij} n_i n_j \int\limits_{-\infty}^{\infty} \exp\left[ -\frac{1}{2} z^2 \left( \frac{1}{h_i^2} + \frac{1}{h_j^2} \right) \right]  \mathrm{d}z \\
        = \quad &\frac{R_{ij}}{\sqrt{2\pi \left( h_i^2+h_j^2 \right)}} N_i N_j \equiv \tilde{R}_{ij} N_i N_j.
    \end{split}
\end{equation}

Vertically integrating the Smoluchowski equation can therefore be achieved by simply replacing the midplane volume densities $n_i$ with the vertically integrated number densities $N_i$ and by multiplying the collision rates with a correction factor. The quantity $\tilde{R}_{ij}$ is stored as \texttt{kernel} in \dustpy{}.

\dustpy{} does not store the vertically integrated number densities but uses dust surface densities, which are simply given by
\begin{equation}
    \Sigma_{\mathrm{d}, i} = m_iN_i.
\end{equation}
Note that by using surface densities instead of number densities, this introduces further mass factors in quantities like $C_{ijk}$, $A^*_i$, or $H_{ijk}$ introduced above. This is straightforward to do, but increases the complexity of the equations presented here. We therefore refer the interested reader to the software for details on the implementation.

Further note, that only the dust densities have been vertically integrated. Other quantities, like the relative velocities needed for $\tilde{R}_{ij}$, are still calculated in the midplane. This is a valid simplification, since most of the mass will settle anyway rather quickly toward the midplane.

\subsection{Algorithm} \label{sec:dust_algo}

\begin{figure}[tb]
    \centering
    \includegraphics[width=\linewidth]{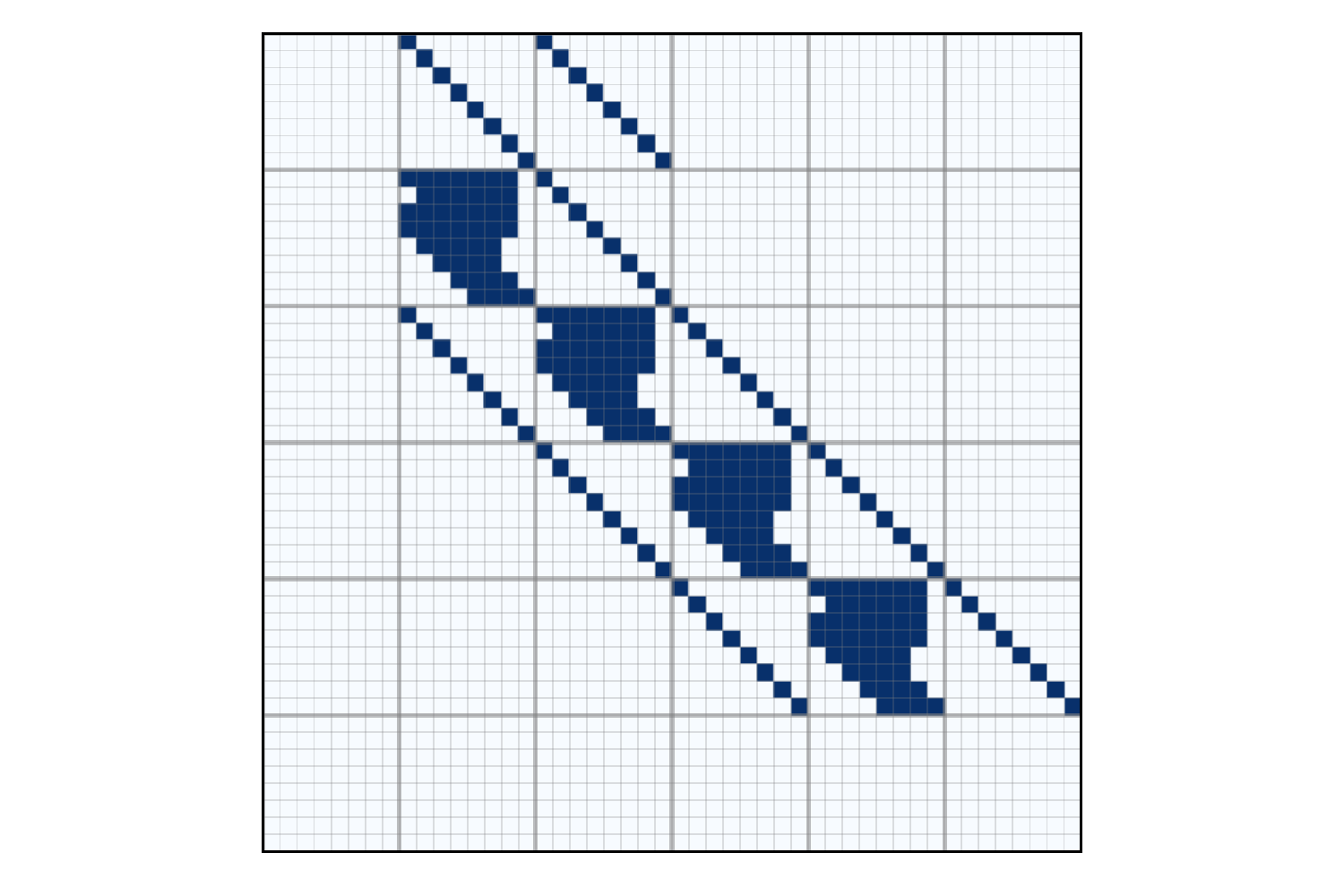}
    \caption{Sketch of the dust Jacobian in an example with six radial grid cells and eight mass bins with only sticking and no fragmentation. The Jacobian has a size $N_rN_m \times N_rN_m$. The $6\times6$ larger squares represent the radial grid, while the smaller $8\times8$ subgrids within each larger square represent the mass grid. Since for dust growth dust particles only collide with dust particles within the same radial grid cell, the Jacobian is only densely filled along the large diagonal squares. For dust transport particles only interact with dust particles of the same mass in the same and adjacent radial grid cells. This is represented in the Jacobian by the main diagonal and the two off-diagonals. The first $N_m$ and last $N_m$ rows are exceptions, since they are used to set the boundary conditions.}
    \label{fig:dust_jacobian}
\end{figure}

Similar to the gas evolution algorithm, dust evolution can be written as a matrix equation. To achieve this, the two dimensional -- distance and mass -- dust surface densities are flattened into a one-dimensional vector
\begin{equation}
    \Sigma_\mathrm{d} \left(r_j, m_k\right)= \Sigma_{\mathrm{d},\left(j-1\right)N_m+k} \equiv \Sigma_{\mathrm{d}, i}.
\end{equation}
With this definition the dust evolution equation can be written in an implicit form
\begin{equation}
    \label{eqn:dust_matrix}
    \begin{split}
        \frac{\vec{\Sigma}_\mathrm{d}^{n+1}-\vec{\Sigma}_\mathrm{d}^n}{\Delta t} &= \left(\mathbb{J}^\mathrm{hyd} + \mathbb{J}^\mathrm{coag}\right)\cdot \vec{\Sigma}_\mathrm{d}^{n+1} + \vec{S}_\mathrm{ext} \\
        &= \mathbb{J} \cdot \vec{\Sigma}_\mathrm{d}^{n+1} + \vec{S}_\mathrm{ext}.
    \end{split}
\end{equation}
The Jacobian in the case of dust evolution consists of two parts, hydrodynamic transport and dust growth. This equation can be solved for the new dust surface densities via
\begin{equation}
    \vec{\Sigma}_\mathrm{d}^{n+1} = \left(\mathbb{1}-\Delta t\mathbb{J}\right)^{-1} \cdot \left( \vec{\Sigma}_\mathrm{d}^n + \Delta t \vec{S}_\mathrm{ext}^n \right)
\end{equation}
by inverting the matrix $\mathbb{1}-\Delta t\mathbb{J}$.

\autoref{fig:dust_jacobian} shows a sketch of the dust Jacobian in the case of six radial grid cells and eight mass bins. The Jacobian has a size of $N_rN_m\times N_rN_m$. The large coarse boxes represent the radial grid cells, while the fine grids within the larger boxes represent the mass grid. As was the case for gas evolution, grid cells only interact with themselves or with neighboring radial grid cells for dust transport. These are the main diagonal and the off-diagonals that are $N_m$ rows above and below the main diagonal.

In the case of dust growth, mass bins can only interact with mass bins in the same radial grid cell. These are the filled boxes along the main diagonal in the Jacobian. The boxes are not completely filled, because in the case of sticking -- that is shown here -- not all collisions are possible. The Jacobian is set to zero for collisions that would result in particles that are larger than the mass grid. Furthermore, the lower-left triangle within a box is empty, because these terms have been added to the respective entries in the upper triangle to save loop iterations in the software.

The first $N_m$ and last $N_m$ rows of the Jacobian are used to set the boundary conditions without calculating coagulation here. In the default \dustpy{} model the inner boundary is set to a constant gradient, while the outer boundary is set to a floor value. Diffusion is turned off at the boundaries by setting the diffusivity to zero.

Since most of the entries of the Jacobian are zero -- in a typical simulation only about $1\,\%$ of the Jacobian is filled -- the Jacobian is stored in a sparse matrix format using \texttt{scipy.sparse}. To invert the matrix and solve the system of equations, in the default \dustpy{} model the matrix is factorized with \texttt{scipy.sparse.linalg.splu}, before the equation is solved with \texttt{scipy.sparse.linalg.SuperLU.solve}. Factorizing the matrix before inversion reduced the runtime of the code in most cases.

Since, the inversion of a large matrix is computationally heavy, \dustpy{} also has two additional options to integrate the dust quantities for large simulation sizes. The first option is to use the generalized minimal residual method, which is an iterative solver, for which \dustpy{} uses \texttt{scipy.sparse.linalg.gmres}. The second option is to integrate the dust quantities explicitly using the fifthth-order adaptive Cash-Karp integration scheme, which does not require the inversion of a matrix.

The timestep $\Delta t$ is calculated such that neither the gas nor the dust densities could become negative in a first-order Euler scheme, while only considering the negative source terms $\dot{\Sigma}_\mathrm{g,\,d}^\mathrm{-}$:
\begin{equation}
    \Delta t = 0.1 \times \min \left| \frac{\Sigma_\mathrm{g,\,d}}{\dot{\Sigma}_\mathrm{g,\,d}^\mathrm{-}} \right|.
\end{equation}

\section{Test Cases} \label{sec:bench}

There are a few test cases with analytical solutions that can be used to benchmark \dustpy{} against. In this section we compare the dust growth algorithm against two collision kernels with analytical solutions and we compare the gas evolution algorithm against the self-similar solutions for viscous accretion.

\subsection{Dust Coagulation} \label{sec:bench_coag}

\begin{figure*}[tb]
    \centering
    \includegraphics[width=0.49\textwidth]{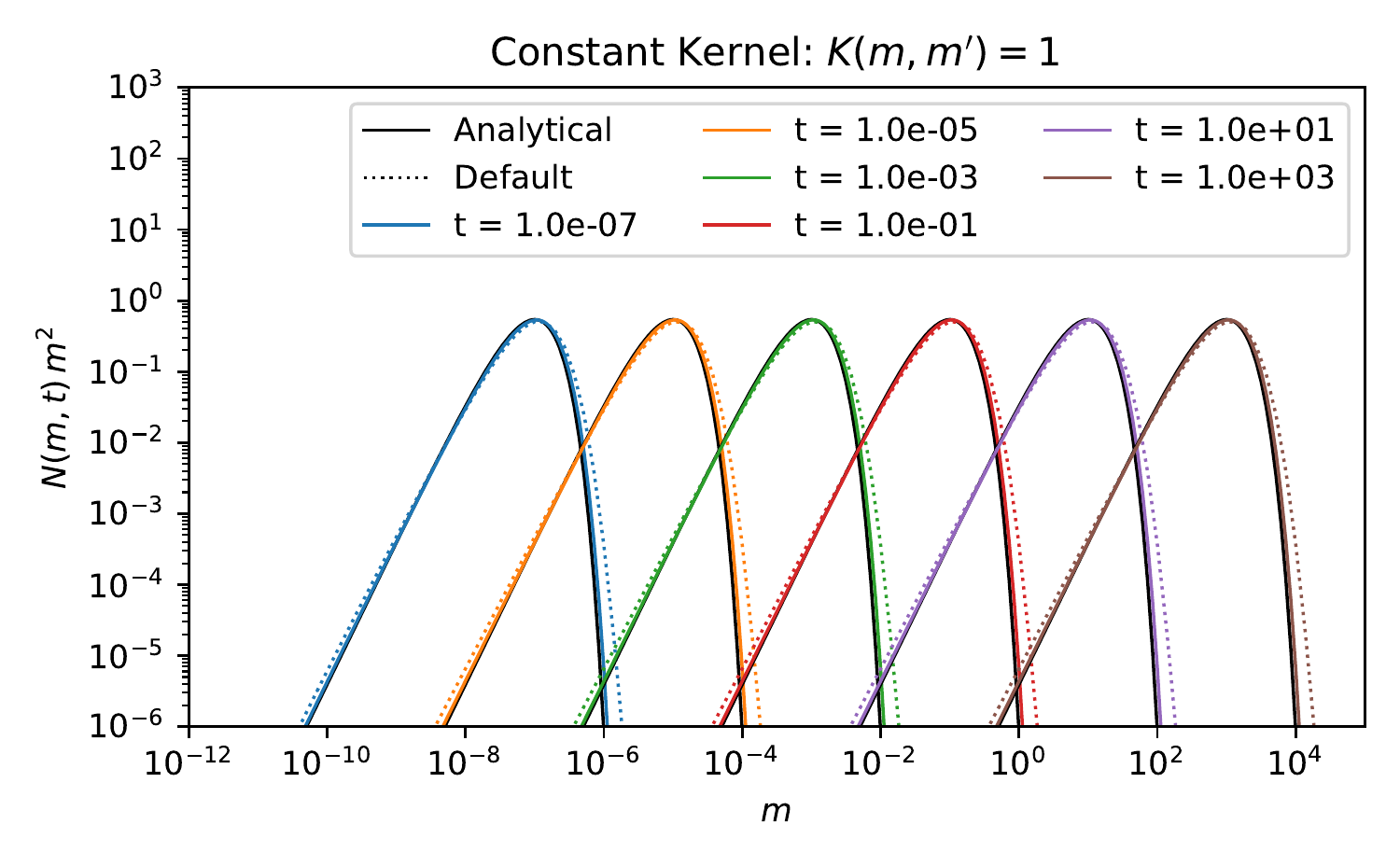}
    \includegraphics[width=0.49\textwidth]{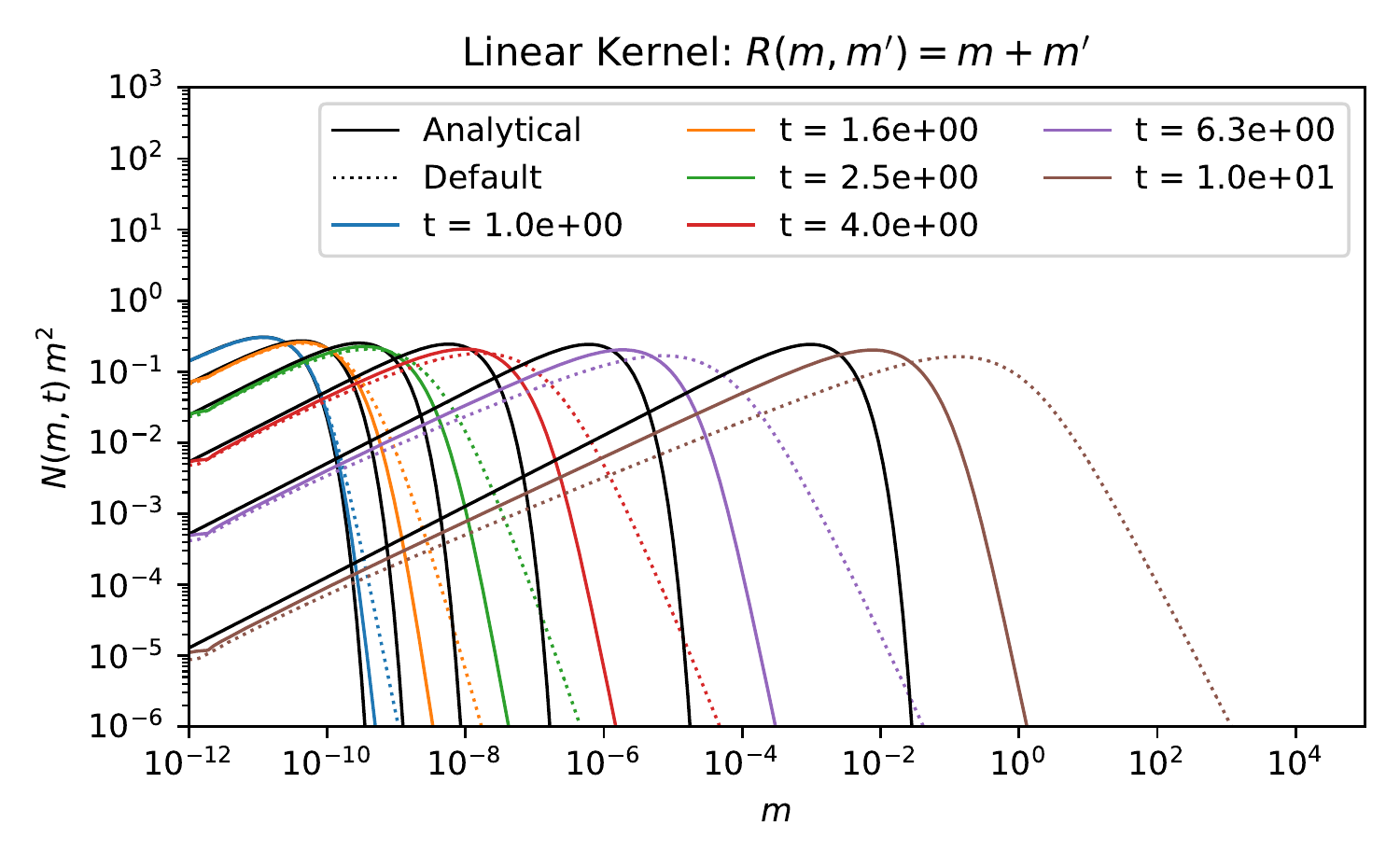}
    \caption{Comparison of the analytical solutions (black solid lines) of the constant kernel (left) and the linear kernel (right) to \dustpy{} simulations with the default mass resolution (dotted colored lines) and with a four times higher mass resolution (solid colored lines).}
    \label{fig:analytical_kernels}
\end{figure*}

Starting from the Smoluchowski equation (\ref{eqn:smolu})
\begin{equation}
    \begin{split}
        \frac{\partial}{\partial t} n\left( m \right) = \int\limits_0^\infty \int\limits_0^{m'} & K\left(m, m', m''\right) R\left(m', m''\right) \quad \times \\
        \times \quad & n\left(m'\right) n\left(m''\right) \mathrm{d}m'' \mathrm{d}m' \\
        -n\left(m\right) \int\limits_0^{\infty} & R\left(m, m'\right)n\left(m'\right)\mathrm{d}m'.
    \end{split}
\end{equation}
and assuming perfect sticking, i.e., $K\left(m, m', m''\right) = \delta\left(m-m'-m''\right)$, leads to
\begin{equation}
    \label{eqn:bench_equation}
    \begin{split}
        \frac{\partial}{\partial t} n\left( m \right) = &\int\limits_0^\infty R\left(m', m-m'\right) n\left(m'\right) n\left(m-m'\right) \mathrm{d}m' \\
        -\quad&n\left(m\right)\int\limits_0^{\infty}R\left(m, m'\right)n\left(m'\right)\mathrm{d}m'.
    \end{split}
\end{equation}
This equation has analytical solutions for three special cases: the constant kernel $R\left(m, m'\right)=\alpha$, the linear kernel $R\left(m, m'\right)=\alpha\left(m+m'\right)$, and the product kernel $R\left(m, m'\right)=\alpha mm'$. The discretized form of this equation for pure sticking was derived in section \ref{sec:dust_coag}.

We will discuss the constant and the linear kernel in this section. The product kernel represents runaway growth that would quickly accumulate the entire mass of the system into a single particle, which cannot be properly addressed within \dustpy{}, as it uses particle distributions instead of physical particles. Solutions to the constant and the linear kernel are discussed in \citet{silk1979ApJ...229..242S} and \citet{wetherill1990Icar...88..336W}.

\subsubsection{The Constant Kernel}

The solution of equation (\ref{eqn:bench_equation}) with the constant kernel is given by
\begin{equation}
    \label{eqn:analytical_constant}
    n\left(m, t\right) = \frac{N_0}{m_0} \left( \frac{2}{\alpha N_0 t} \right)^2 \exp \left[ \frac{2}{\alpha N_0 t} \left( 1 - \frac{m}{m_0} \right) \right],
\end{equation}
where $m_0$ is the smallest possible mass and $N_0$ the initial total number density of particles:
\begin{equation}
    N_0 = \int\limits_0^\infty n\left(m, 0\right)\mathrm{d}m.
\end{equation}
Note that the \dustpy{} code units are the number densities integrated over the mass bin. We therefore initialize the simulation by setting the first mass bin to $N_0$ and all other mass bins to zero.

The result is shown in \autoref{fig:analytical_kernels} (left panel) for a simulation with the default mass resolution of \dustpy{} with dotted lines and for a simulation with a four times higher mass resolution with solid lines for $\alpha=1$. The analytical solutions given by equation (\ref{eqn:analytical_constant}) are plotted with black solid lines. There is a slight deviation visible at the upper mass tail of the distribution in the default resolution run. An explanation for this is given at the end of the next section.

\subsubsection{The Linear Kernel}

The solution of the linear kernel is given by
\begin{equation}
    \label{eqn:analytical_linear}
    \begin{split}
        n\left(m, t\right) = &\frac{N_0}{2\sqrt{\pi}m_0^2} \frac{g}{\left(1-g\right)^{0.75}} \times \\
        \times &\exp \left[ -\frac{m}{m_0} \left( 1 - \sqrt{1-g} \right)^2 \right],
    \end{split}
\end{equation}
with $N_0$ being the initial total number density as for the constant kernel and
\begin{equation}
    g = \exp \left[ -\alpha N_0 m_0 t \right].
\end{equation}
Initially, only the first mass bin was filled with $N_0$ while all other mass bins were set to zero.

The result is shown in \autoref{fig:analytical_kernels} (right panel) for a simulation with the default mass resolution of \dustpy{} with dotted lines and for a simulation with a four times higher mass resolution with solid lines for $\alpha=1$. The analytical solutions given by equation (\ref{eqn:analytical_linear}) are plotted with black solid lines. As for the constant kernel there is a deviation at the upper mass end of the distribution that is worse the lower the mass resolution is.

The reason for this is the algorithm described in section \ref{sec:dust_coag}. Since the mass grid is logarithmically spaced, the combined mass of both colliding particles in a sticking collision will not directly fall onto the mass grid itself, but has to be distributed between the adjacent mass bins as described in equation (\ref{eqn:lin_cont}). This leads to artificial growth, because material will be added to a bin that is more massive than the combined mass of the collision partners. This causes the simulations to be more massive at the higher mass end and -- due to mass conservation -- less massive at the lower mass end compared to the analytic solutions. The situation is worse for the linear kernel, because the kernel is proportional to the colliding mass itself. An overestimation of mass will overestimate the kernel itself.

But since the computational time is highly sensitive to the number of mass bins, one has to find a compromise between accuracy and execution time. For the simple collision model in the default \dustpy{} simulation, the default mass resolution should be sufficient, since growth will be eventually halted by the fragmentation or by the drift barrier. Only the growth time scale might be slightly underestimated. For more complex collision models, including, for example, mass transfer, a higher mass resolution might be crucial. For more details on this we refer to \citet{drazkowska2014A&A...567A..38D}, which performed mass resolution tests for more complex collision models. In any case, we advise to always run selected simulations with a higher mass resolution to verify that the default mass resolution was sufficient.

\begin{figure}[tb]
    \centering
    \includegraphics[width=\linewidth]{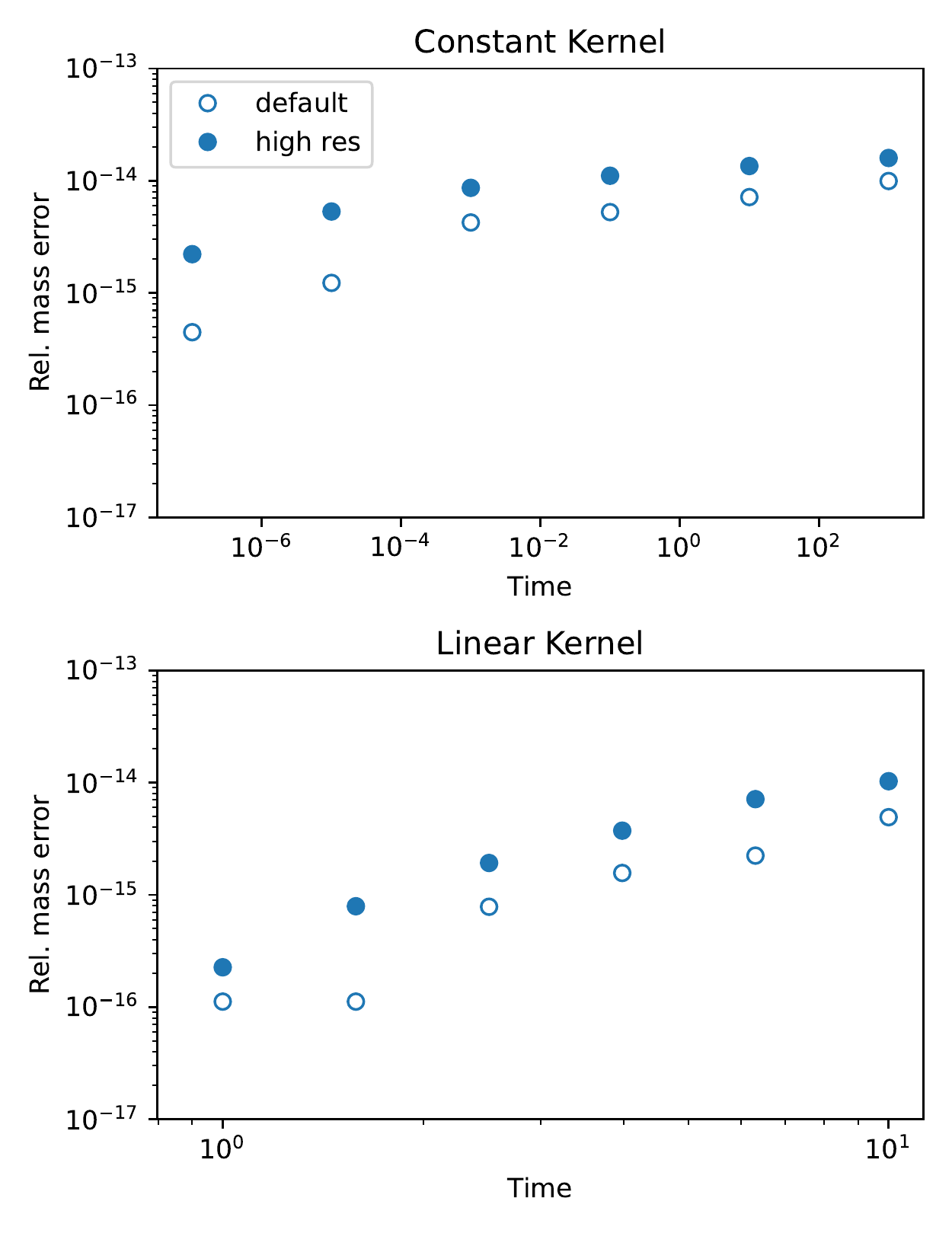}
    \caption{Relative error in mass in the benchmarks for the constant (top) and linear (bottom) kernels in the default (open circles) and the high-resolution (filled circles) runs.}
    \label{fig:mass_error}
\end{figure}

\autoref{fig:mass_error} shows the relative errors in mass for the two benchmark models of the constant and linear kernels with the default and the high-resolution runs. In all cases the errors are very close to machine-precision levels for double-precision floating point numbers. The coagulation algorithm of \dustpy{} is therefore mass conserving.

\subsection{Gas Transport} \label{sec:bench_trans}

\begin{figure}[tb]
    \centering
    \includegraphics[width=\linewidth]{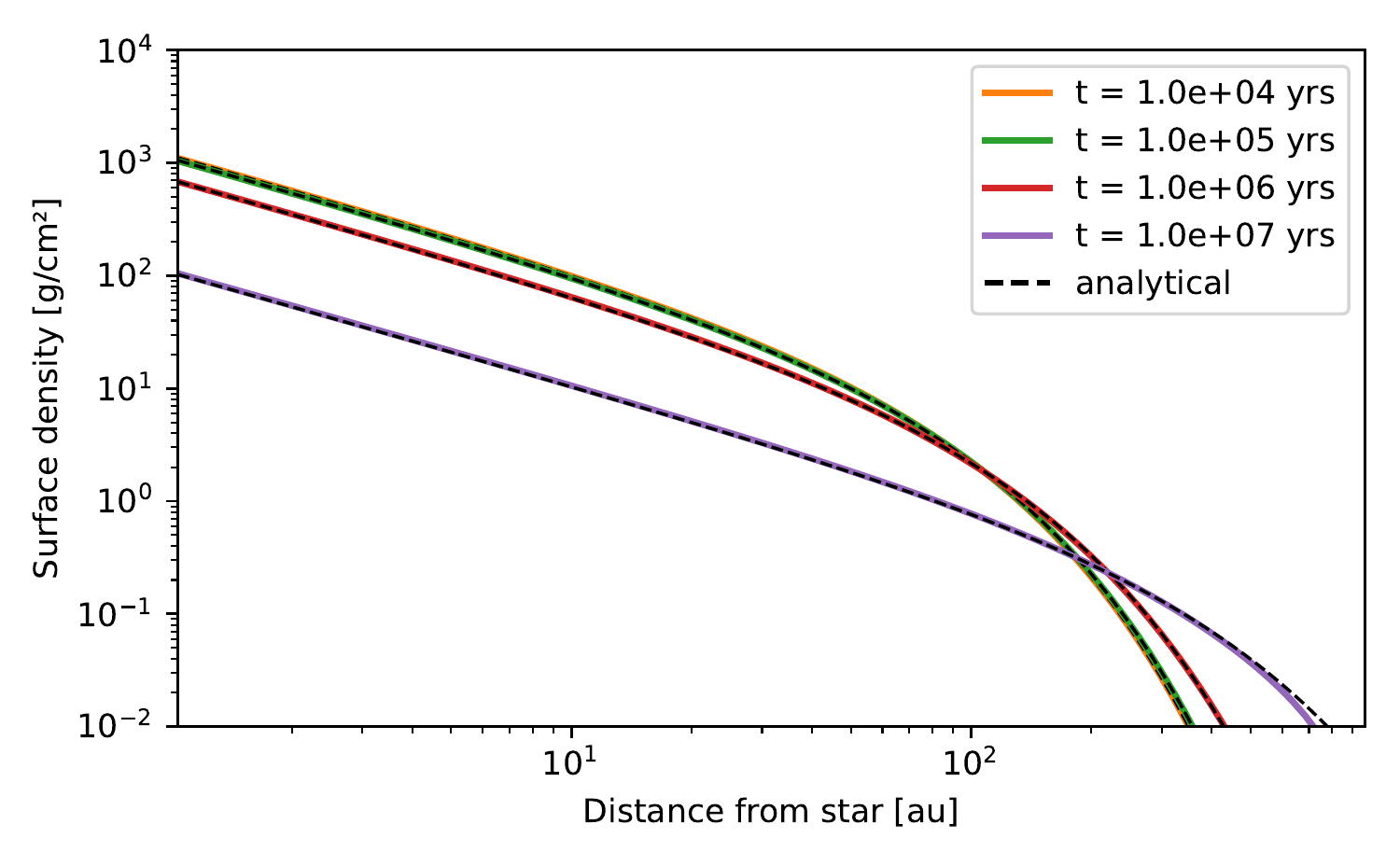}
    \caption{Comparison of the gas evolution of \dustpy{} against the analytical solution of equation (\ref{eqn:gas_evo_anal}).}
    \label{fig:gas_benchmark}
\end{figure}

Viscous gas accretion as given by equations (\ref{eqn:gas_evo}) and (\ref{eqn:gas_vvisc}) has an analytical solution, as discussed by \citet{lynden-bell1974MNRAS.168..603L} and \citet{hartmann1998ApJ...495..385H} and is given by
\begin{equation}
    \label{eqn:gas_evo_anal}
    \Sigma_\mathrm{g} \left(R\right) = \frac{C}{3\pi\nu_1R^\gamma} T^{-\frac{5/2-\gamma}{2-\gamma}} \exp \left( -\frac{R^{2-\gamma}}{T} \right),
\end{equation}
with the dimensionless time $T=\frac{t}{t_\mathrm{s}} + 1$, with a dimensionless scaling of the radial grid $R=\frac{r}{r_1}$, and with the viscosity at the scaling location $\nu_1 = \nu\left(r_1\right)$. $\gamma$ is the exponent of the viscosity assuming it is a power law:
\begin{equation}
    \label{eqn:gamma_nu}
    \nu \left( r \right) = \alpha \cdot c_\mathrm{s}^2 \cdot \frac{1}{\Omega_\mathrm{K}} \propto r^s \cdot r^q \cdot r^\frac{3}{2} \equiv r^\gamma.
\end{equation}
In the default \dustpy{} simulation with $s=0$ and $q=-\frac{1}{2}$ it follows that $\gamma=1$. The scaling factor of time is given by
\begin{equation}
    t_\mathrm{s} = \frac{1}{3\left(2-\gamma\right)^2}\frac{r_1^2}{\nu_1}
\end{equation}
and the mass normalization factor
\begin{equation}
    C = M_0 \frac{3\nu_1}{2r_1^2\left(2-\gamma\right)},
\end{equation}
with $M_0$ being the initial mass of the disk. We compare in \autoref{fig:gas_benchmark} a simulation with the default \dustpy{} parameters against the analytical solution given by equation (\ref{eqn:gas_evo_anal}). As can be seen, the result of \dustpy{} is in good agreement with the analytical solution. Only in the last snapshot there is a small deviation close to the outer edge of the grid. The reason for this is that the outer gas boundary is set to the gas floor value, which is a very small number. This causes a minor underestimation of the surface density when the disk expands and reaches the outer boundary.

A word of caution on other slopes of the surface density: with the above default parameters of \dustpy{} the surface density slope in the inner disk will be $-1$. If one wants to achieve slopes other than $-1$, it is not enough to simply change the initial surface density profile. Over time the surface density will approach $-\gamma$ as given by equation (\ref{eqn:gamma_nu}), since the viscosity profile determines the gas profile in the long run. One has to change the slopes of the viscosity or the temperature profile accordingly.

Similarly, the inner boundary has to be changed from a constant gradient to constant power law, for any other surface density profile than $-1$. For more details we refer to the documentation\footnote{Documentation: \href{https://stammler.github.io/dustpy/}{https://stammler.github.io/dustpy/}}.

\section{Examples} \label{sec:examples}

\begin{figure*}[tb]
    \centering
    \includegraphics[width=\textwidth]{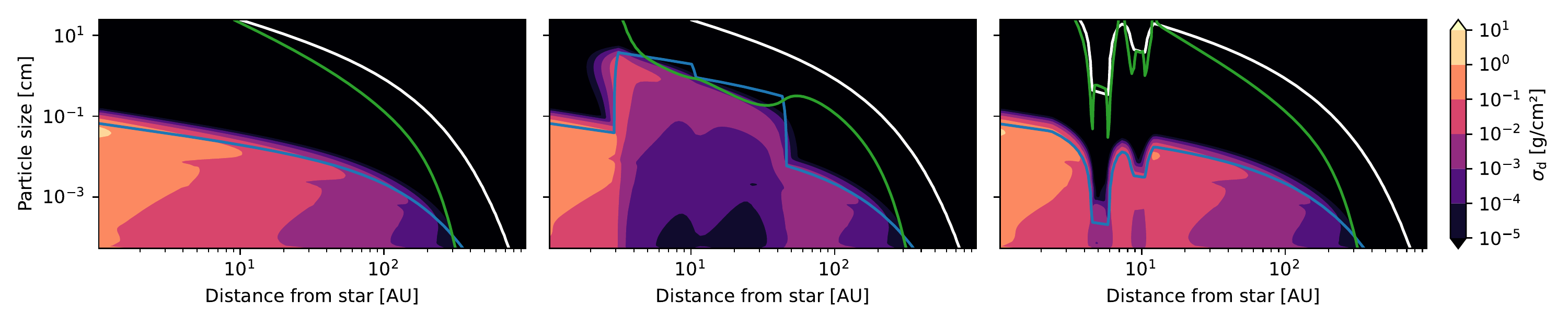}
    \caption{Example simulations of dust evolution with \dustpy{} after a simulation time of $1\,\mathrm{Myr}$. Left: the default model. Center: the default model with three ice lines changing the fragmentation velocity. Right: the default model including Jupiter and Saturn opening gaps.}
    \label{fig:examples}
\end{figure*}

In this section we discuss selected examples of simulations performed with \dustpy{}. The left panel of \autoref{fig:examples} shows the default model of \dustpy{} that is run when no parameter or function has been modified. The plotted snapshot is after $1\,\mathrm{Myr}$. The quantity $\sigma_\mathrm{d}$ that is plotted is defined by
\begin{equation}
    \Sigma_\mathrm{d} \left( r \right) = \int\limits_0^\infty \sigma_\mathrm{d}\left(r, m\right) \mathrm{d} \log m,
\end{equation}
such that it is independent of the mass grid, since the code units of \dustpy{} are integrated over the mass bin and therefore depend on the mass grid.

Table \ref{tab:parameters} lists the key parameters of the default model. But since \dustpy{} is under continuous development, these parameters might be subject to change in the future. We would therefore like to refer to the documentation\footnote{Documentation: \href{https://stammler.github.io/dustpy/}{https://stammler.github.io/dustpy/}}. This will always list the most recent model parameters.
The default temperature profile is that of a passively irradiated disk with an irradiation angle of $0.05$
\begin{equation}
    \label{eqn:gas_temp}
    T\left(r\right) \approx \frac{T_*}{2} \sqrt{\frac{R_*}{r}} \propto r^{-\frac{1}{2}}
\end{equation}
with the stellar radius and temperature $R_*$ and $T_*$.

\begin{deluxetable*}{clcc}
    \label{tab:parameters}
    \tablecaption{Key parameters of the default \dustpy{} model.}
    \tablewidth{0pt}
    \tablehead{
        \colhead{Parameter} & \colhead{Description} & \colhead{Value} & \colhead{Equation}
    }
    \startdata
    $R_\mathrm{in}$ & Inner grid boundary & $1\ \mathrm{AU}$ & \\
    $R_\mathrm{out}$ & Outer grid boundary & $1000\ \mathrm{AU}$ & \\
    $N_r$ & Number of radial grid cells & $100$ \\
    \hline
    $m_\mathrm{min}$ & Minimum particle mass & $10^{-12}\ \mathrm{g}$ & \\
    $m_\mathrm{max}$ & Maximum particle mass & $10^{5}\ \mathrm{g}$ & \\
    $N_\mathrm{mbpd}$ & Number of mass bins per decade & $7$ & \\
    \hline
    $M_*$ & Stellar mass & $1\ M_\odot$  & \\
    $R_*$ & Stellar radius & $2\ R_\odot$  & (\ref{eqn:gas_temp}) \\
    $T_*$ & Stellar effective surface temperature & $5772\ \mathrm{K}$  & (\ref{eqn:gas_temp}) \\
    \hline
    $M_\mathrm{disk}$ & Initial disk mass & $0.05\ M_\odot$  & \\
    $p$ & Power law of surface density & -1 & \\
    $R_\mathrm{c}$ & Initial critical cut-off radius & $30\ \mathrm{AU}$ & \\
    & Initial dust-to-gas ratio & $10^{-2}$  & \\
    \hline
    $\alpha$ & $\alpha$-viscosity parameter & $10^{-3}$ & (\ref{eqn:gas_vvisc}) \\
    $\delta_\mathrm{r}$ & Radial mixing parameter & $10^{-3}$ & (\ref{eqn:dust_diff}) \\
    $\delta_\mathrm{t}$ & Turbulent mixing parameter & $10^{-3}$ & \\
    $\delta_\mathrm{z}$ & Vertical mixing parameter & $10^{-3}$ & (\ref{eqn:dust_h}) \\
    \hline
    $a_\mathrm{max}^\mathrm{ini}$ & Maximum initial particle size & $1\ \mu\mathrm{m}$ & \\
    $\beta$ & Initial particle size distribution $n\left(a\right)\propto a^{\beta}$ & $-3.5$ & \\
    $\gamma$ & Fragment distribution & $-11/6$ & (\ref{eqn:fragment_dist}) \\
    & Mass ratio for erosion & $10$ & \\
    $\rho_\mathrm{s}$ & Dust bulk mass density & $1.67\ \mathrm{g\,cm^{-3}}$ & \\
    $v_\mathrm{frag}$ & Fragmentation velocity & $1\ \mathrm{m\,s^{-1}}$ & (\ref{eqn:p_frag}) \\
    $\chi$ & Excavated erosive mass fraction & $1$ & (\ref{eqn:mass_normalization}) \\
    \enddata
\end{deluxetable*}

The blue line in \autoref{fig:examples} is the fragmentation barrier. As particles grow their relative velocities increase as shown in \autoref{fig:rel_vel}. If their relative velocities exceed the fragmentation velocity, which is $1\,\frac{\mathrm{m}}{\mathrm{s}}$ in the default model, the particles start to fragment instead of growing further. The fragmentation barrier is an estimate by \citet{birnstiel2012A&A...539A.148B} and is given by
\begin{equation}
    a_\mathrm{frag} = \frac{2}{3\pi} \frac{\Sigma_\mathrm{g}}{\rho_\mathrm{s}\delta_t} \frac{v_\mathrm{frag}^2}{c_\mathrm{s}^2},
\end{equation}
where $a_\mathrm{frag}$ is the maximum size a particle can reach at any location in the disk where particle growth is fragmentation limited.

The green line is the drift barrier. As seen in equation (\ref{eqn:dust_v_rad}), particles have increasing drift speeds with increasing Stokes numbers, i.e., with increasing size, until they reach the maximum drift speed at a Stokes number of unity. At some point the particles drift more rapidly toward the star, before they can grow to larger sizes. This is called drift barrier and estimated by \citet{birnstiel2012A&A...539A.148B} as
\begin{equation}
    a_\mathrm{drift} = \frac{2}{\pi} \frac{\Sigma_\mathrm{d}}{\rho_\mathrm{s}} \left(\frac{H_\mathrm{P}}{r}\right)^{-2} \left| \frac{\partial \log P}{\partial \log r} \right|^{-1},
\end{equation}
where $a_\mathrm{drift}$ is the maximum size particles can reach anywhere in the disk where growth is drift limited. The white lines in \autoref{fig:examples}, represent particle sizes with a Stokes number of unity, i.e., particles with the highest drift speeds and highest relative velocities. To achieve this simulation result only seven lines of \texttt{Python} code were needed including the import of modules and the initialization.

The center panel of \autoref{fig:examples} shows the default model but with three ice lines at which the fragmentation velocity is changing. The model is similar to the model of \citet{pinilla2017ApJ...845...68P}. The idea behind the model is that the fragmentation velocity of particles depends on their chemical composition, where icy particles could be more sticky than pure silicate particles \citep{schaefer2007A&A...470..733S, wada2009ApJ...702.1490W}. The adopted fragmentation velocity in this model is given by
\begin{equation}
    v_\mathrm{frag} =
    \begin{cases}
        1\,\frac{\mathrm{m}}{\mathrm{s}}  & \text{for $T>150\,\mathrm{K}$ (pure silicate)}            \\
        10\,\frac{\mathrm{m}}{\mathrm{s}} & \text{for $150\,\mathrm{K}>T>80\,\mathrm{K}$ (water ice)} \\
        7\,\frac{\mathrm{m}}{\mathrm{s}}  & \text{for $80\,\mathrm{K}>T>40\,\mathrm{K}$ (ammonia)}    \\
        1\,\frac{\mathrm{m}}{\mathrm{s}}  & \text{else (carbon dioxide)}
    \end{cases}
\end{equation}
Please note, however, that newer experiments suggest that the fragmentation velocity does not change that dramatically with composition \citep{musiolik2019ApJ...873...58M}. In regions with higher fragmentation velocity particles can grow to larger sizes, before they fragment. Since larger particles drift more rapidly, this depletes the outer disk rather quickly, enriching the inner disk inside the water ice line with material. To set up this model, only 14 lines of \texttt{Python} code were required, mainly to write a function for the fragmentation velocity. Please note further, that this example is only a showcase of the potential of \dustpy{} and does not include evaporation and condensation of the molecules in question. It is relatively easy to include additional gas species in \dustpy{}. The addition of new dust parameters like ice contents would require greater modifications to the dust distributions \citep[see][]{okuzumi2009ApJ...707.1247O, stammler2017A&A...600A.140S}.

The right panel of \autoref{fig:examples} shows the default model but with Jupiter and Saturn inserted at their current location. The planets open gaps in the gas disk and, since dust particles follow pressure gradients, the gap is also cleared of dust. It is difficult to achieve this gap opening in an one-dimensional simulation self-consistently. We therefore use the gap profiles provided by \citet{kanagawa2017PASJ...69...97K} obtained from two-dimensional hydrodynamical simulations. Since we also want to model gas accretion, we cannot simply set the gas surface density directly to these profiles. But since the product of the viscosity $\nu$ and the gas surface density $\Sigma_\mathrm{g}$ is a constant in steady state, we can impose the inverse of the gap profile on the viscosity.

As seen in \autoref{fig:examples}, the planetary regions are cleared of gas. Since the Stokes number is inversely proportional to the gas surface density, the white line for particles of Stokes number unity is directly proportional to the gas surface density. The particles in the planetary regions have larger Stokes numbers due to the reduced gas surface density leading to higher relative velocities and therefore smaller particles sizes. To set up this simulation about 80 lines of \texttt{Python} code were necessary, where most of the code is needed to define the gap profiles of \citet{kanagawa2017PASJ...69...97K}.

Further examples of research that have been done with \dustpy{} include \citet{stammler2019ApJ...884L...5S}, which implemented planetesimal formation in dust rings at the outer edges of gaps to explain the observed optical depths in protoplanetary disks. In a similar model, \citet{miller2021MNRAS.tmp.2674M} showed that moving pressure bumps could explain the observations of wide exo-Kuiper belts. \citet{garate2020A&A...635A.149G} implemented back reaction of dust particles onto the gas in \dustpy{} to investigate the influence of dust enrichment at ice lines on the accretion rate. \citet{pinilla2021A&A...645A..70P} performed a parameter study on the $\alpha$, $\delta_r$, $\delta_t$, and $\delta_z$ parameters to investigate their influence on the maximum particle sizes dust particles can reach. \citet{drazkowska2021A&A...647A..15D} used \dustpy{} to benchmark a simple model for the prediction of the pebble accretion rate in protoplanetary disks.

\section{Summary}
\label{sec:summary}

We developed \dustpy{}, a \texttt{Python} package to simulate dust evolution in protoplanetary disks. \dustpy{} solves for viscous gas evolution, dust advection and diffusion, and dust growth by coagulation and fragmentation. Computationally expensive routines are written in \texttt{Fortran} and called from within the \texttt{Python} environment.

\dustpy{} uses the \texttt{Simframe} framework for scientific simulations, which makes it easy to change every aspect of the code to allow for a multitude of research opportunities.

Dust growth is simulated by solving the Smoluchowski equation for a dust mass distribution. In contrast to Monte Carlo methods, which simulate the evolution of representative particles, the advantage of \dustpy{} is its execution time. The default model of \dustpy{} takes about 25\,minutes to evolve a full protoplanetary disk for 100,000 years. The Monte Carlo model of \citet{drazkowska2013A&A...556A..37D}, for example, takes about 25\,days to simulate an annulus within a protoplanetary disk for 30,000 years with 100,000 representative particles. The advantage of Monte Carlo methods, however, is that it is straightforward to add additional parameters to them, such as porosity or ice fraction to the dust particles, while this requires greater modifications to the dust densities in Smoluchowski solvers \citep[see][]{okuzumi2009ApJ...707.1247O, stammler2017A&A...600A.140S}.

\dustpy{} is a one-dimensional code that only has one spatial coordinate, the radial distance from the star, to simulate axisymmetric disks. Nonaxisymmetric features as reported by \citet{drazkowska2021A&A...647A..15D} can therefore not be modeled. The vertical extent of the disk -- the height above the midplane -- is assumed to be always in hydrostatic equilibrium. This assumption is good enough for most parts of the disk, but might be violated in parts, where the collisional timescale becomes significantly shorter than the mixing timescale \citep[see][]{krijt2016ApJ...822..111K, klarmann2018A&A...618L...1K}. Furthermore, sedimentation-driven coagulation by particles settling toward the midplane cannot be modeled \citep{Zsom2011}.

\dustpy{} uses a logarithmic mass grid to cover large dynamic ranges from submicron interstellar medium grain sizes to meter-sized boulders. Particles resulting from hit-and-stick collisions will therefore not generally lie on the mass grid itself. Their mass will be added into the two adjacent mass bins. This will, however, artificially create particles that are too large. In the default model this will not be an issue since dust growth is halted by fragmentation and drift; only the growth timescale will be slightly underestimated. In more complex collision model, as in \citet{windmark2012A&A...544L..16W}, where large particles can continue growing by sweeping up small particles, it is crucial to not overestimate the sizes of the largest particles. Other algorithms, like \citet{lee2000Icar..143...74L}, \citet{dullemond2005A&A...434..971D}, or \citet{lombart2021MNRAS.501.4298L}, may be better suited to conserve the shape of the particle distribution. The coagulation algorithm of \dustpy{}, on the other hand, conserves the dust mass up to machine precision. In any case, we advise users to always compare their results to high-resolution runs and check for convergence.

\begin{acknowledgments}

    We thank the referee for their very helpful comments and suggestions as well as the rigorous cross-checking of the equations presented in this publication.
    The authors acknowledge funding from the European Research Council (ERC) under the European Union's Horizon 2020 research and innovation program under grant agreement No. 714769 and funding by the Deutsche Forschungsgemeinschaft (DFG, German Research Foundation) under grant Nos. 361140270, 325594231, and under Germany's Excellence Strategy - EXC-2094 - 390783311.

\end{acknowledgments}


\bibliography{bibliography}{}
\bibliographystyle{aasjournal}



\end{document}